\documentclass[preprint,showpacs,preprintnumbers,amsmath,amssymb,nofootinbib,floatfix,tightenlines]{revtex4}

\usepackage{graphicx}
\usepackage{dcolumn}
\usepackage{bm}

\begin{document}
 

\title{Self-consistent description of multipole strength: systematic 
calculations}

\author{J.~Terasaki}
 \affiliation{School of Physics, Peking University, Beijing 100871, P.~R.~China}
\author{J.~Engel}
 \affiliation{Department of Physics and Astronomy,\\
University of North Carolina, Chapel Hill, NC 27599-3255}

\date{\today}

\begin{abstract}  

We use the quasiparticle random phase approximation with a few   Skyrme density
functionals to calculate strength functions in the $J^\pi = 0^+, 1^-,$ and
$2^+$ channels for even Ca, Ni, and Sn isotopes, from the proton drip line to
the neutron drip line.  We show where and how low-lying strength begins to
appear as $N$ increases.  We also exhibit partial energy-weighted sums of the
transition strength as functions of $N$ for all nuclei calculated, and
transition densities for many of the interesting peaks.

We find that low-energy strength increases with $N$ in all multipoles, but with
distinctive features in each. 
The low-lying $0^+$ strength near the neutron at large $N$ barely involves 
protons at
all, with the strength coming primarily from a single two-quasineutron
configuration with very large spatial extent.  The low-lying $1^-$ strength is
different, with protons contributing to the transition density in the nuclear
interior together with neutrons at large radii.
The low-lying $2^+$ transition strength goes largely to more localized states.
The three Skyrme interactions we test produce similar results, differing most
significantly in their predictions for the location of the neutron drip line,
the boundaries of deformed regions, energies of and transition strengths to the
lowest $2^+$ states between closed shells, and isovector energy-weighted sum
rules.

\end{abstract}

\maketitle


\section{\label{sec:introduction}Introduction}

Very unstable nuclei with unusual ratios of neutron number ($N$) to proton number
($Z$) are currently the subject of intense theoretical and experimental work.  A
particularly important issue is the effect on nuclear properties of a large
neutron excess and the accompanying low-lying continuum of excited states.
In ground states these conditions sometimes lead to halos, thick skins, and
changes in magic
numbers.  Less is known about excited states, which we study here by
calculating transition strengths and densities.

Most of the work on excited states is in light neutron-rich nuclei; a
low-energy strength-function peak has been studied for years.  Recent work
\cite{Kob89,Ber90,Bertr92}  has focused on enhanced $1^-$
strength in nuclei like $^{11}$Li or in neutron-rich oxygen isotopes
\cite{Toh94,Yok95,Cat97,Ham98,Ham99,Vre01,Col01,Mat05,Sag99}.   But in
heavier nuclei, calculations are more spotty.  Several groups have calculated
the isovector (IV) $1^-$ strength functions, or photoabsorption cross sections,
in the Ca isotopes \cite{Suz90,Cham94,Cat97,Sag01,Mat05}.  References
\cite{Cat97,Sag01} predict a small peak at around $E=9$ MeV in $^{60}$Ca.  The
first paper also calculates the strength function in $^{48}$Ca, finding no
low-energy peak, but  Ref.~\cite{Cham94}, a calculation based on 
density-functional theory, finds a very broad bump at $E=6$--10 MeV.  It is not
clear, therefore, at what value of $N$ the low-energy strength is first
discernible. 

The method we use here, the quasiparticle random phase approximation (QRPA),
has been applied mainly to selected $1^-$
excitations in particular isotopes of O, Ca, and Ni;  one example is Ref.\
\cite{Mat05}, which asserts that the low-energy excitations reflect
strong particle-particle (``dineutron") correlations. 
References \cite{Gor03,Gor04} show extensive and systematic QRPA calculations of
photoabsorption cross sections, but focus primarily on radiative
neutron-capture cross sections.  The relativistic
QRPA \cite{Vre01,Paa03,Vre04,Paa05} has been applied more with an eye to nuclear
structure issues (for a recent review, see Ref.~\cite{Vre05}), asserting, for
example, that low-lying isoscalar (IS) $1^-$ states exhibit toroidal flow.
References \cite{Sar04,Tso04} go beyond RPA, altering details of predicted low-lying
strength.

Other channels have been discussed less than the IV $1^-$.  The authors of
Ref.~\cite{Ham97} calculate $0^+$ strength functions in Ca isotopes up to the
neutron drip line, finding an enhancement at low energy in $^{60}$Ca. 
Reference \cite{Ham96} examines transition strengths in the $0^+$, $1^-$, and $2^+$
channels near the neutron drip line, finding, for example, a sharp
$2^+$ peak near $E=0$ in $^{110}$Ni.

The above represents a considerable amount of work on excitations in exotic
nuclei.  Its range and results can be summarized, roughly, as follows: 
1) Most of the calculations are for spherical nuclei, and
the IV $1^-$ channel has been studied the most.
2) A low-energy peak grows as $N$ increases.
3) The low-energy excitations do not seem collective  
in the $1^-$ channel (e.g.~Ref.~\cite{Col03}), 
i.e.\ they do not contain more than a few
particle-hole or two-quasiparticle configurations.
4) The corresponding transition densities have long 
neutron tails extending beyond the bulk nuclear radius. 
5) The neutron skin thickness is correlated with
the amount of low-energy strength \cite{Pie06}.
6) Calculations reproduce measurements of the amount of low-energy
strength fairly well if they include many-particle many-hole correlations 
\cite{Sar04,Tso04}.

Most of this work has focused on nuclei right at the drip line and, as already
noted, has
emphasized the IV $1^-$ channel.  It is still not known whether theory makes
similar predictions for other neutron-rich nuclei, or to what extent the
properties of the low-energy peaks vary from one multipole to another.  The
purpose of this paper, therefore, is to calculate strength functions in as many
nuclei as possible, in all multipoles up to $2^+$, and investigate how the
properties of the strength functions vary as $N$ and the multipolarity change.
We use
the QRPA with Skyrme density functionals (which we refer to as
interactions from here on) and volume-type delta pairing interactions.  

Currently, the QRPA is the most sophisticated method that can be used
systematically for resonances of all energies in
heavy nuclei without assuming that part of the nucleus forms an inert core. 
Here we will display results of QRPA calculations for Ca, Ni,
and Sn isotopes from the proton drip line to the neutron drip line, describing
the corresponding strength functions and transition densities in detail.  We
will also examine whether results depend on which Skyrme interaction we use, so
that we can identify important measurements/nuclei that will help distinguish
among them.

Section \ref{sec:calculation} below describes our method. 
Section \ref{sec:strength function} shows strength functions
in the three isotopic chains and looks for
$(N,Z)$-dependent trends in their behavior. 
Section \ref{sec:transition density}
discusses transition densities to states in low-lying peaks.
Section \ref{sec:conclusion} is a conclusion.

\section{\label{sec:calculation}Calculation}

Our approach to the QRPA is described in detail in Ref.\ \cite{Ter05}, and
we discuss it here only briefly. 
The starting point is a coordinate-space Hartree-Fock-Bogoliubov (HFB)
calculation, carried out
with the code used in Refs.~\cite{Dob84,Dob96}.  The code outputs
quasiparticle wave functions, with spherical symmetry assumed.  It takes the continuum into account by placing the  nucleus in a 20-fm box.  We use as large
a single-particle space as possible and solve the HFB equations accurately, so
as to ease the removal of spurious states from the QRPA and satisfy sum rules.

We diagonalize the HFB density matrix to obtain a set of canonical states
--- as many as there are quasiparticle states ---
that make up our QRPA basis.
The QRPA-Hamiltonian matrix then contains off-diagonal elements from the
quasiparticle one-body Hamiltonian as well as the matrix elements of the residual
two-body interaction.  The two-body matrix elements are functional 
derivatives of the energy with respect to the density and pairing tensor,
After diagonalizing the QRPA Hamiltonian, we use the eigenvectors to obtain
strength functions \cite{Rin80} and transition densities (see Appendix). 

We cannot handle an infinite number of canonical quasiparticle states,
and so truncate our basis in the following way:
For $N < 82$ the maximum 
single-particle angular momentum $j_{\rm max}$ is 15/2,
and for $N \geq 84$ $j_{\rm max}$ is 21/2. 
If the ground state has a finite pairing gap in the HFB calculation 
we truncate the canonical quasiparticle basis in the QRPA by omitting states
with very small 
occupation probability $v_i^2$. In the $J^\pi = 0^+$ and $1^-$ channel we
keep states with $v_i^2 \geq 10^{-12}$, and in the $J^\pi = 2^+$ channel we keep those
with $v_i^2 \geq 10^{-8}$.
If the mean field has no pairing we use the Hartree-Fock energy to truncate; in
the $J^\pi = 0^+$ and $1^-$ channels we keep single-particle 
states with energy $\varepsilon_i \leq 150$ MeV, and in the $J^\pi = 2^+$
channel we keep those with $\varepsilon_i \leq 100$ MeV.  In the HFB
calculations we always cut off the quasiparticle spectrum at 200
MeV.

The largest QRPA-Hamiltonian matrix that we encounter here (in the $1^-$
channel for Sn near the neutron drip line) is about 14,000, 
and the minimum is about 2,000 (in the $0^+$ channel
for Ca near the proton drip line).\footnote{In Ref.~\cite{Ter05}, cut-offs of
$v_i^2 = 10^{-12}$ and $\varepsilon_i = 150$ MeV were used for the $2^+$
channel, and then the maximum dimension of the Hamiltonian matrix was about
20,000.}
The separation of spurious excitations from
physical states is as complete as in Ref.\ \cite{Ter05} in most of cases.

We use 3 Skyrme interactions: SkM$^\ast$, SLy4, and SkP. 
Our pairing interaction has zero range with no density
dependence (this is ``volume pairing''). 
For reasons we do not understand, we cannot obtain solutions of the HFB equation with SkP for Ca and Ni isotopes. 
The other interactions give solutions that are unstable against QRPA quadrupole vibrations
for isotopes with:
$N=72$ in Ni (SkM$^\ast$);  
$N=104-112$ (SkM$^\ast$) and
$N=60-64$, $96-110$ (SLy4) in Sn, 
and so we do not discuss these nuclei.  

\section{\label{sec:strength function} Strength functions} 
Strength functions are important because they capture information about all excited states and can often
be measured.  We define the strength functions through the transition operators

\noindent
$J^\pi=0^+$:
\begin{eqnarray} 
&&\hat{F}_{00}^{\rm IS}=\frac{eZ}{A}\sum_{i=1}^Ar_i^2,\nonumber\\
&&\hat{F}_{00}^{\rm IV}=\frac{eN}{A}\sum_{i=1}^Zr_i^2
-\frac{eZ}{A}\sum_{i=1}^Nr_i^2,
\end{eqnarray}

\noindent
$J^\pi=1^-$:
\begin{eqnarray} 
&&\hat{F}_{1M}^{\rm IS}=\frac{eZ}{A}\sum_{i=1}^Ar_i^3\,Y_{1M}(\Omega_i),
\nonumber\\
&&\hat{F}_{1M}^{\rm IV}=\frac{eN}{A}\sum_{i=1}^Zr_i\,Y_{1M}(\Omega_i)
-\frac{eZ}{A}\sum_{i=1}^Nr_i\,Y_{1M}(\Omega_i),
\end{eqnarray}

\noindent
$J^\pi=2^+$:
\begin{eqnarray} 
&&\hat{F}_{2M}^{\rm IS}=\frac{eZ}{A}\sum_{i=1}^Ar_i^2\,Y_{2M}(\Omega_i),
\nonumber\\
&&\hat{F}_{2M}^{\rm IV}=\frac{eN}{A}\sum_{i=1}^Zr_i^2\,Y_{2M}(\Omega_i)
-\frac{eZ}{A}\sum_{i=1}^Nr_i^2\,Y_{2M}(\Omega_i),
\label{eq:transition_operator_2+}
\end{eqnarray}
where $A$ is total-nucleon number, and 
$i$ is a single-particle index.
The upper limit for $Z$ ($N$) on a summation sign means that the sum is over
protons (neutrons). 
 
In the next few sections we show the strength functions, with spurious states
removed, produced by SkM$^\ast$,
and discuss those produced by the other Skyrme interactions when appropriate.  We then display 
energy-weighted sums (up to several different energies) produced by all our 
interactions, in all isotopes. 

\subsection{\label{subsec:strength_function_0+} The 0$^{\bm  +}$ channels} 

\begin{figure}
\includegraphics[width=1.0\textwidth]{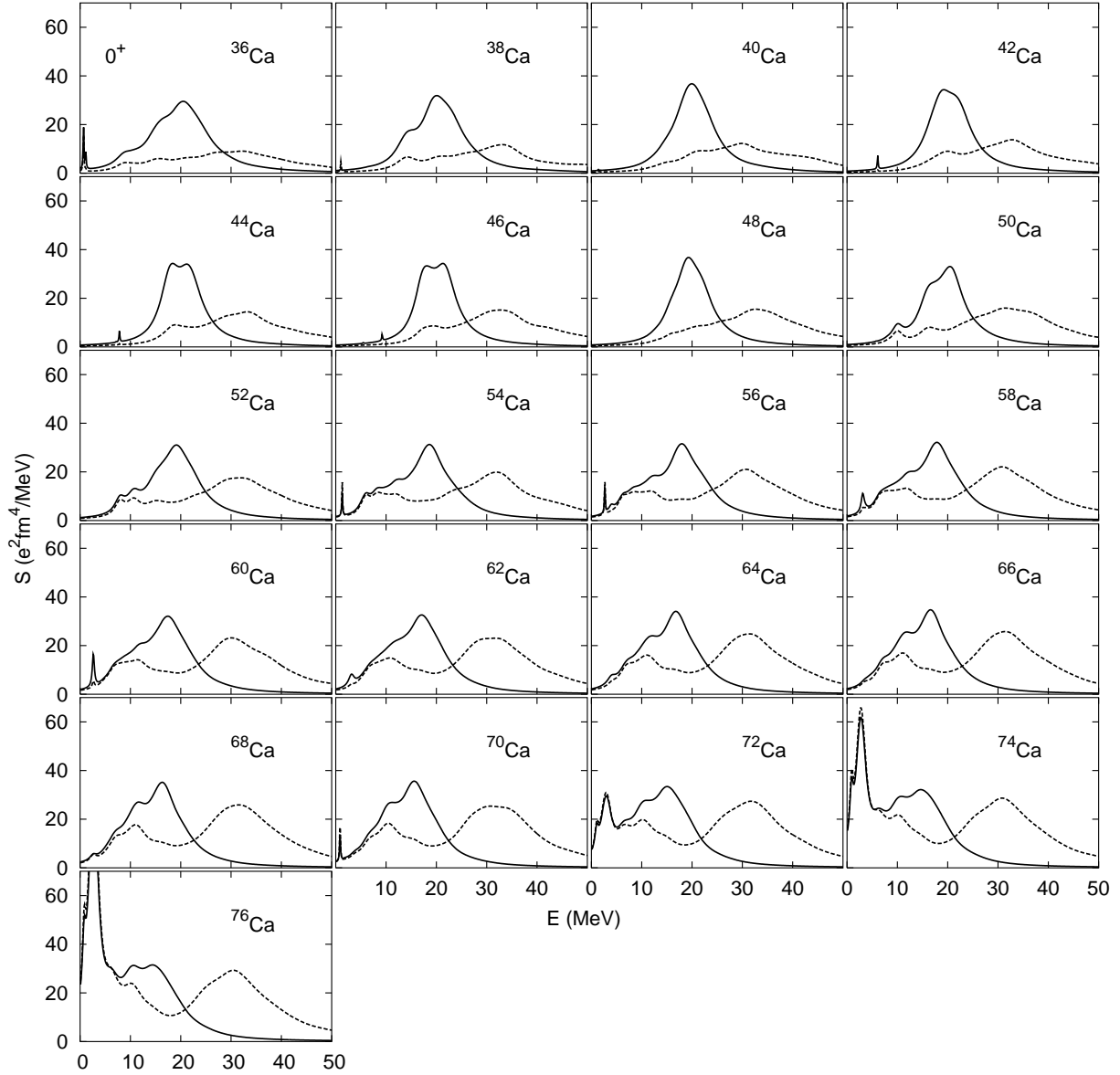}
\caption{\label{fig:str_ca_0+} 
IS (solid) and IV (dashed)
0$^+$ strength functions for even Ca isotopes (SkM$^\ast$).}
\end{figure}

\begin{figure} 
\includegraphics[width=1.0\textwidth]{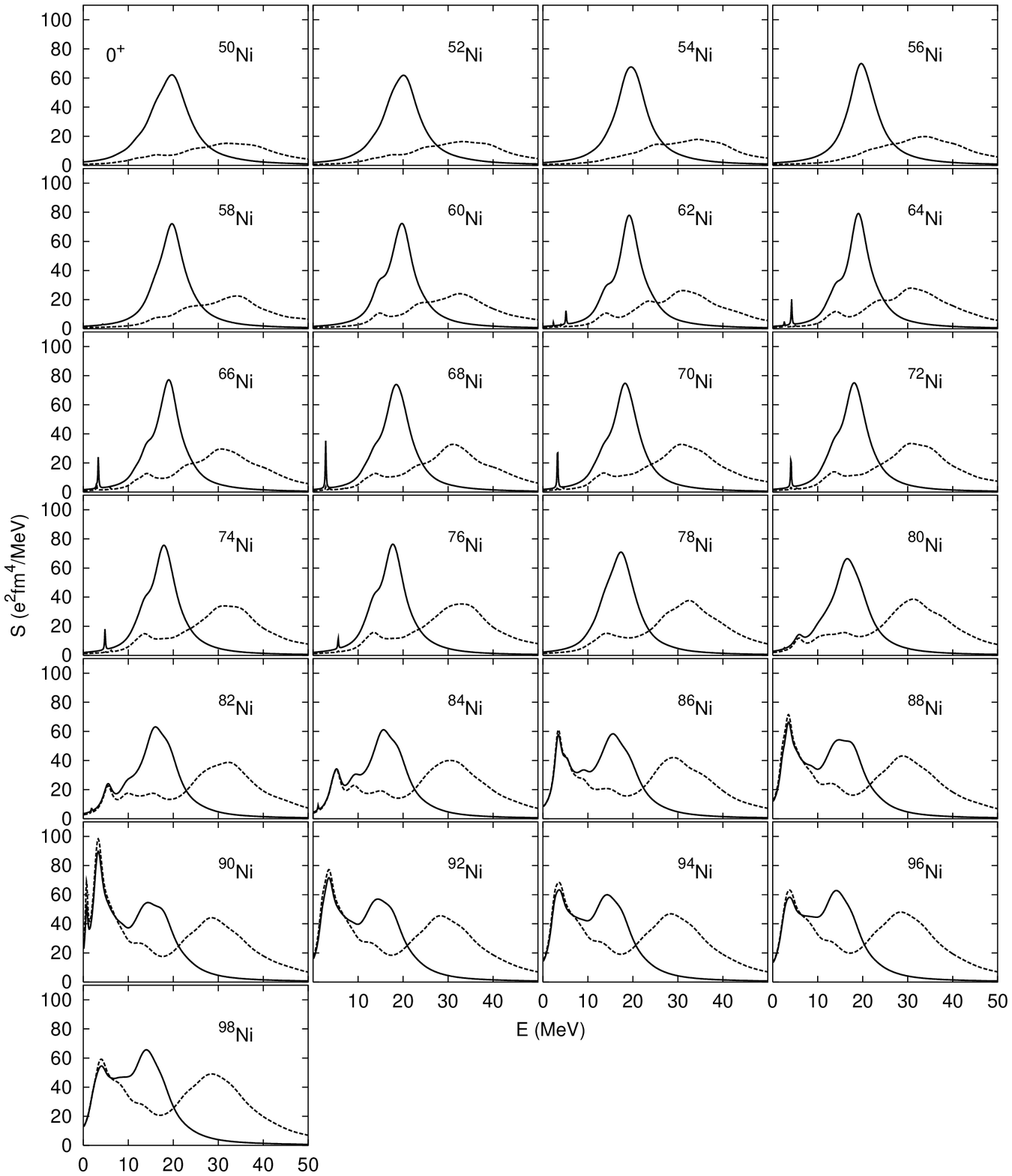}
\caption{\label{fig:str_ni_0+} 
The same as Fig.~\ref{fig:str_ca_0+} but for Ni isotopes. }
\end{figure}

\begin{figure}
\includegraphics[width=1.0\textwidth]{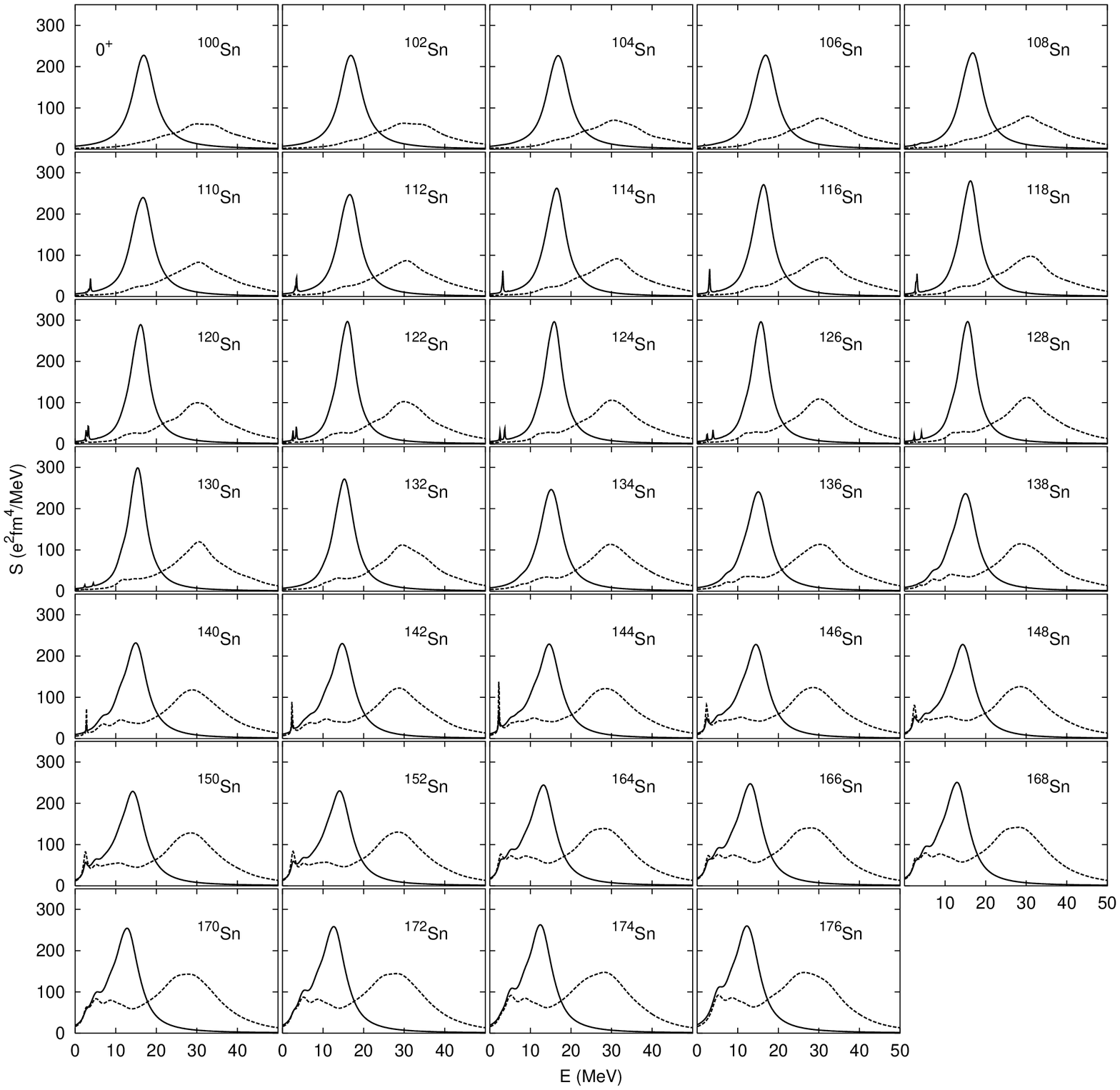}
\caption{\label{fig:str_sn_0+} 
The same as Fig.~\ref{fig:str_ca_0+} but for Sn isotopes. }
\end{figure}

Figures \ref{fig:str_ca_0+}--\ref{fig:str_sn_0+} show the IS and IV strength
functions in the $J^\pi = 0^+$ channels.  In Ca (Fig.\
\ref{fig:str_ca_0+}), most of the functions are smooth except for small
low-energy spikes in several nuclei and the large low-energy peaks in $^{74,76}$Ca.  The IS giant resonances peak 
at $E=20$ MeV for $^{36-52}$Ca and gradually move to 15 MeV by $^{76}$Ca. 
As $N$
approaches the neutron drip line, a peak near $E=0$ grows, becoming quite large
at the drip line.  The IV giant resonance is broad in the light isotopes and
develops a low-energy component as $N$ increases, apparently increasing the
summed strength.  The larger peak varies only slightly in energy over the same
range of $N$.  Near the drip line the low-energy IV strength develops a peak
that exactly mirrors the IS peak, indicating that the strength in both channels is
produced solely by neutrons.  This distinct low-energy bump seems to appear
first at 10 MeV in $^{50}$Ca.  This nucleus is not as short-lived as the
isotopes near the drip line; perhaps the appearance of the low-energy mode is
experimentally testable.  The same nucleus marks the beginning of the low
energy peak when we use SLy4.

The $N$ dependence of the strength function of Ni (Fig.~\ref{fig:str_ni_0+}) is
similar in many ways to that of Ca; the energy of the IS giant-resonance energy,
for example, falls in a similar way past $N =40$.  A large low-energy peak
again develops near the neutron-drip line, though not as dramatically as in
Ca, and with the complication that the strength shrinks after $^{90}$Ni.  This
peak first is noticeable in both the IS and IV channel at 6 MeV in $^{80}$Ni,
here and in calculations with SLy4.
This
example and that in Ca suggest more generally that the nucleus with 2 neutrons
outside a closed shell will mark the appearance of the low-energy peak.


In Sn (Fig.~\ref{fig:str_sn_0+}), the peak near zero energy around the drip
line is much smaller than in Ca and Ni.
The bump first appears in $^{136}$Sn at $E=8$ MeV (violating by two nucleons
the conjecture made a few lines up) in both the 
IS and IV channels.  This threshold nucleus is the same with other Skyrme
interactions.

\begin{figure}
\includegraphics[width=0.8\textwidth]{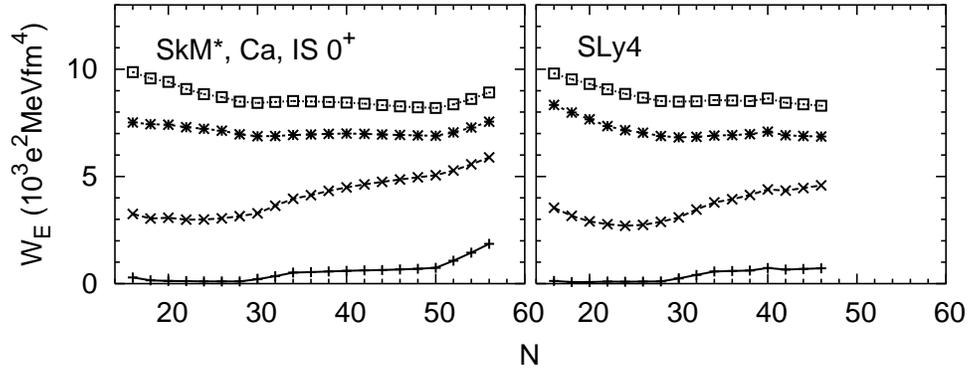}
\caption{\label{fig:part_esum_is_ca0+} 
$N$-dependence of the partial energy-weighted sums in the IS 
$0^+$ channel of Ca.
The upper limits of the sums are 
10 MeV (plus symbol), 20 MeV (x), 30 MeV (asterisk), and 
100 MeV (open square). 
 }
\end{figure}

Although we can make these kinds of comparisons in words, we would
like to display the results of both (or all three where possible)
Skyrme interactions in a more
quantitative and graphic way.  To limit the number of our figures, we
do so by showing predictions for the sum of energy weighted strength up
to 4 successively higher energies.  We define the ``partial'' sum, up to an
energy $E$ as
\begin{equation}
W_E = \int_0^{E}dE'\,E'S(E'),
\label{eq:part_esum}
\end{equation}
where $S(E')$ is a strength function\footnote{If the strength function has a 
large peak near $E=0$, Eq.~(\ref{eq:part_esum}) overestimates the
energy-weighted sum, because the tail in the negative energy 
region is not included. This problem occurs near the neutron drip line, 
and the energy-weighted sum is then overestimated by 10 \% at most.
We did not calculate $W_E$ from negative $E^\prime$, because 
then $W_{10 {\rm MeV}}$ is sometimes negative.
} (see Eq.~(1) of
Ref.~\cite{Ter05}).
Figure \ref{fig:part_esum_is_ca0+} shows $W_E$ with $E=10$, 20, 30, and 100 
MeV for the IS $0^+$ channel in Ca.
All Skyrme interactions show similar $N$-dependence and $E$-dependence; 
the difference between the $E=30$ and 100 MeV curves
is comparable with or smaller than the differences among the other curves 
(separated by 10 MeV), reflecting the eventual vanishing of the 
strength as $E$ increases.
The  $E=20$ MeV curve increases more rapidly with $N$ than the others in 
the region 
$N > 30$, indicating that the strength 
around $E=20$ MeV is shifted down at large $N$.

The differences among the interactions are most apparent in the location
of the neutron drip line, a static property. 
The differences in strength functions, at least in this channel,
are for the most part minor (though large enough to make
significant differences in, e.g, the compression modulus). 

\begin{figure}
\includegraphics[width=0.8\textwidth]{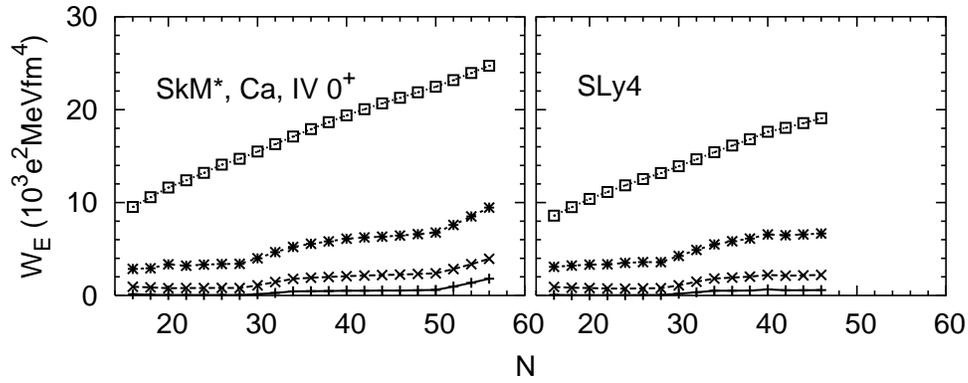}
\caption{\label{fig:part_esum_iv_ca0+} 
The same as Fig.~\ref{fig:part_esum_is_ca0+} but for the 
IV channel. The open squares now correspond to an upper limit 
of 150 MeV. 
 }
\end{figure}

Figure \ref{fig:part_esum_iv_ca0+} shows the partial sums in the IV channel.
Here the two interactions do yield somewhat different results.  Though the
$N$-dependence is similar, SkM$^\ast$ produces more integrated strength than SLy4. 
To see how the total energy-weighted strength
can differ from one interaction to another, we examine the IV 
energy-weighted sum rule (EWSR)
\begin{equation} 
\sum_k\sum_M E_k|\langle k|\hat{F}_{JM}^{\rm IV}|0\rangle|^2
= S_{\rm EW}(J) + C_{\rm c} I_{\rm c}(J),
\label{eq:EWSR_iv}
\end{equation}
\begin{equation}
S_{\rm EW}(0) = \frac{2e^2\hbar^2}{m}\left\{
\frac{N^2Z}{A^2}\langle r^2\rangle_{\rm p} + 
\frac{Z^2N}{A^2}\langle r^2 \rangle_{\rm n}\right\},
\end{equation}
\begin{equation}
S_{\rm EW}(J\geq 1) = \frac{e^2\hbar^2}{8\pi m}J(2J+1)^2
\left\{
\frac{N^2Z}{A^2}\langle r^{2J-2}\rangle_{\rm p} +
\frac{Z^2N}{A^2}\langle r^{2J-2}\rangle_{\rm n}
\right\},
\end{equation}
\begin{equation}
C_{\rm c} = \frac{1}{4}(t_1+t_2)
+\frac{1}{8}(t_1 x_1 + t_2 x_2), \label{eq:C_c}
\end{equation} 
\begin{equation} 
I_{\rm c}(J) =
e^2\sum_M\int d^3\bm{r}\,|\bm{\nabla}f_{JM}(\bm{r})|^2
\rho_0^{\rm n}(\bm{r})
\rho_0^{\rm p}(\bm{r}),
\label{eq:9}
\end{equation}
\begin{eqnarray}
&& f_{00}(\bm{r}) = r^2, \\
&& f_{JM}(\bm{r}) = r^J Y_{JM}(\Omega), \ J \geq 1. 
\end{eqnarray}
Here $E_k$ is the energy of the excited state $|k\rangle$,
$|0\rangle$ is the ground state, $m$ is the nucleon mass, and
$\langle r^2\rangle_{\rm p}$
$(\langle r^2\rangle_{\rm n})$ is the mean square 
proton (neutron) ground-state radius. The sum over states $k$
does not include spurious states.
The second term $C_{\rm c}I_{\rm c}(J)$ in Eq.~(\ref{eq:EWSR_iv}), with $\rho_0^{\rm p}(\bm{r})$ and
$\rho_0^{\rm n}(\bm{r})$ of Eq.\ (\ref{eq:9})
the proton and neutron ground-state densities,
is a correction arising from the momentum-dependent terms of
the Skyrme interaction.
The Skyrme parameters $t_0, t_1,\cdots$ are defined in
Refs.~\cite{Bar82} (SkM$^\ast$), \cite{Cha98} (SLy4), and 
\cite{Dob84} (SkP).  
Another term in the sum rule
depending on the proton-neutron mixed density is always 
zero in our work.
 
In Tab.~\ref{tab:ewsr_iv_ca0+} we show
$S_{\rm EW}(0)$, 
$C_{\rm c}$, $I_{\rm c}(0)$, and EWSR (the full energy-weighted sum) 
for $^{60}$Ca.
The interactions reproduce ground-state observables about equally well, so the
difference in the EWSR, as the table shows, comes from the term involving $C_{\rm
c}$.  The degree by which the EWSR exceeds  $S_{\rm EW}(J)$ is in line
with expectations (see, e.g., Chap.~10 of \cite{Bor98} and
Refs.~\cite{Cha98,Col04}).

\begin{figure} 
\includegraphics[width=0.8\textwidth]{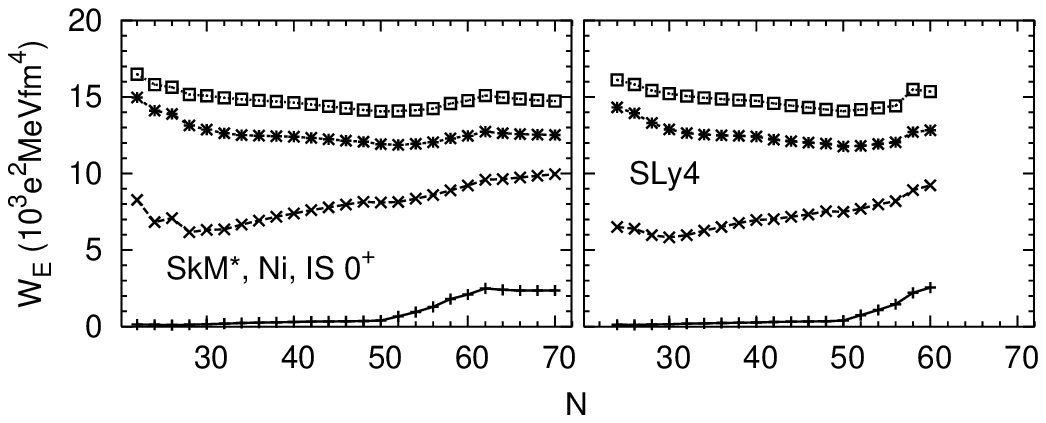}
\caption{\label{fig:part_esum_is_ni0+} 
The same as Fig.~\ref{fig:part_esum_is_ca0+} but for Ni. 
 }
\end{figure}
\begin{figure}  
\includegraphics[width=0.8\textwidth]{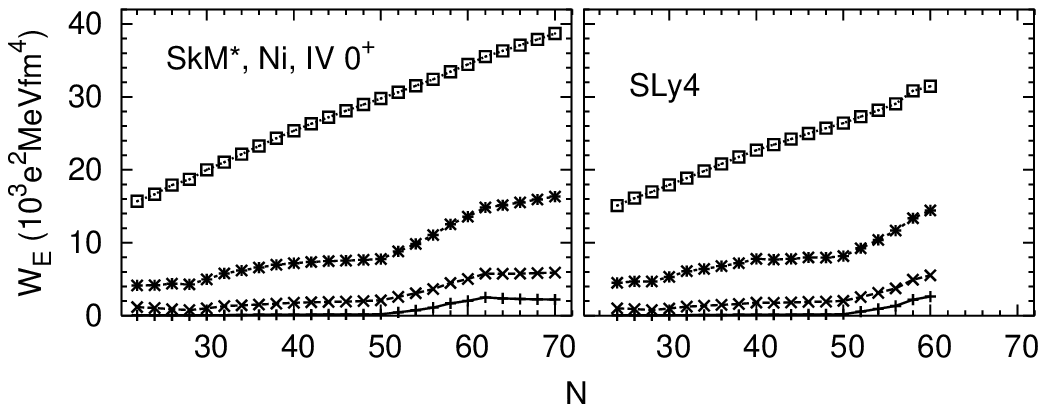}
\caption{\label{fig:part_esum_iv_ni0+} 
The same as Fig.~\ref{fig:part_esum_iv_ca0+} but for Ni. 
 }
\end{figure}
\begin{figure} 
\includegraphics[width=1.0\textwidth]{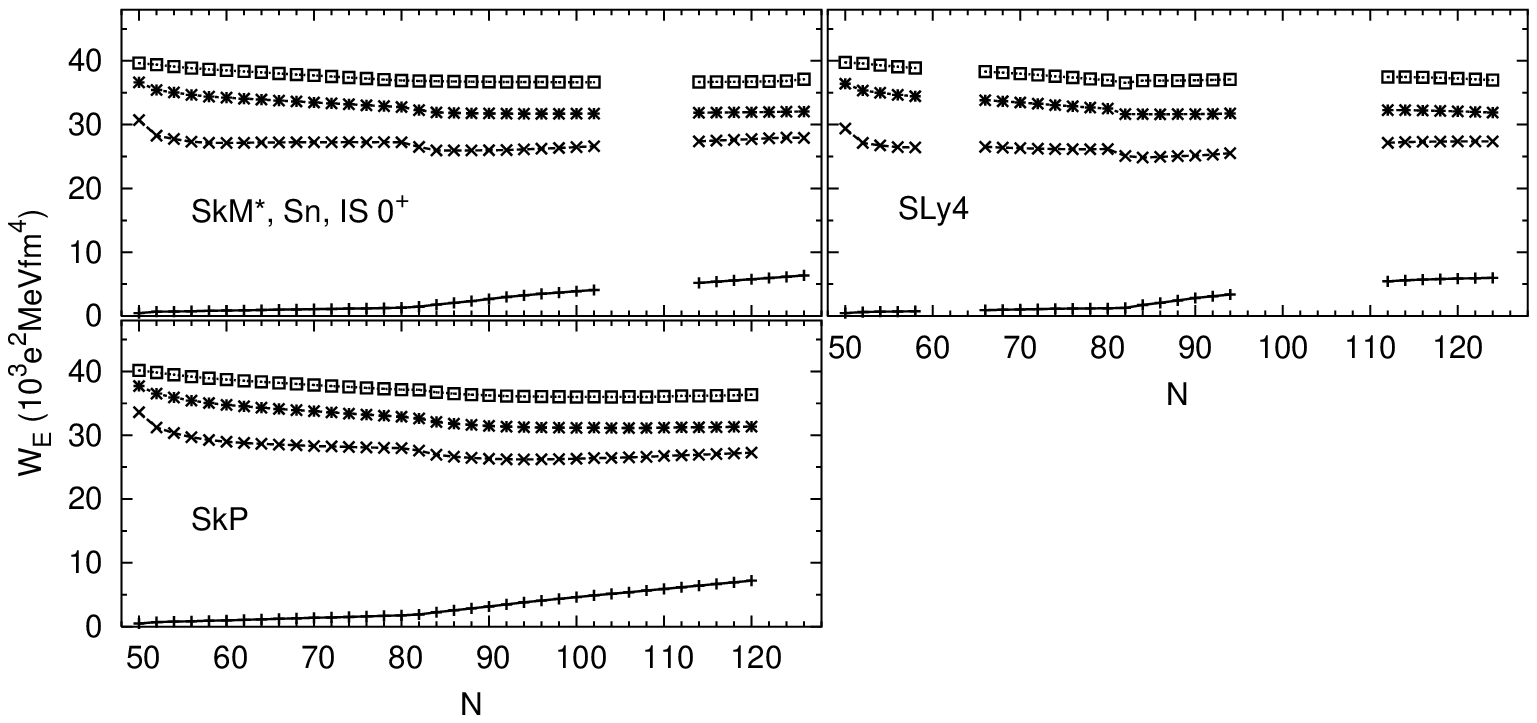}
\caption{\label{fig:part_esum_is_sn0+} 
The same as Fig.~\ref{fig:part_esum_is_ca0+} but for Sn.
 }
\end{figure}
\begin{figure} 
\includegraphics[width=1.0\textwidth]{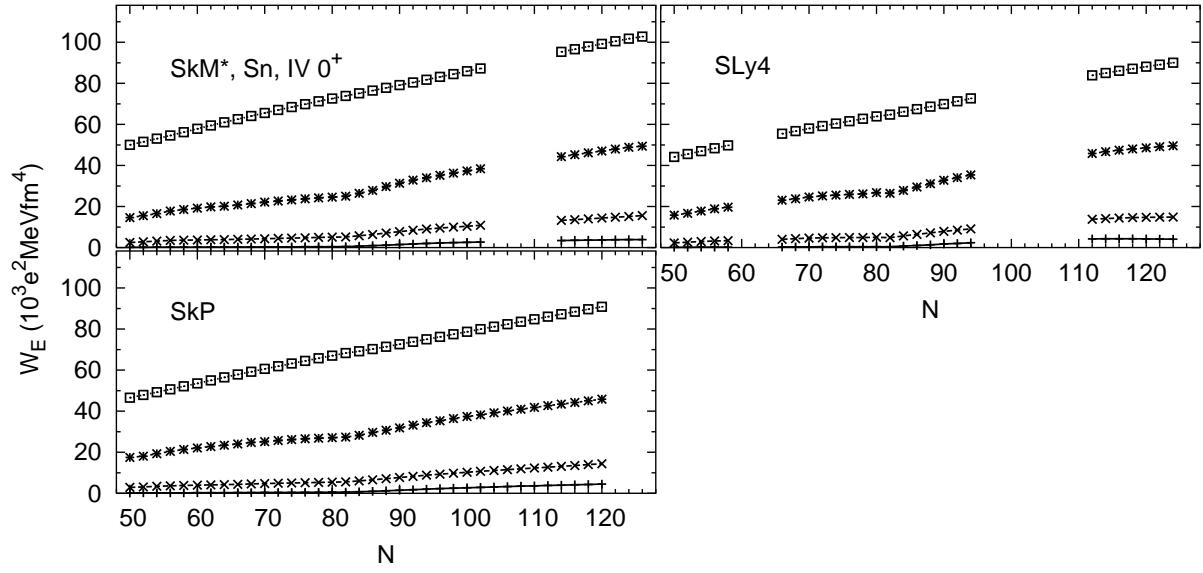}
\caption{\label{fig:part_esum_iv_sn0+} 
The same as Fig.~\ref{fig:part_esum_iv_ca0+} but for Sn.
 }
\end{figure}

The properties of Ni are similar to those of Ca.  
$W_{E=20{\rm MeV}}$ in the IS channel of Ni increases more rapidly than 
the other curves (Fig.~\ref{fig:part_esum_is_ni0+}).
SkM$^\ast$ again has the larger IV sum-rule value.
 (Fig.~\ref{fig:part_esum_iv_ni0+}). 
$W_{E=20{\rm MeV}}$ in the IS channel of Sn is much higher than $W_{E=10{\rm MeV}}$ 
(Fig.~\ref{fig:part_esum_is_sn0+}),
reflecting the lower energy of the giant resonance. 
Figure \ref{fig:part_esum_iv_sn0+} shows that the IV sum rules follow 
the same order as in the other isotopes, with SkP slightly larger
than SLy4.
\clearpage

\begin{table}
\caption{\label{tab:ewsr_iv_ca0+}
Components of the EWSR of IV $0^+$ channel of $^{60}$Ca.
}
\begin{ruledtabular}
\begin{tabular}{ccccc}
parameter set & $S_{\rm EW}(0)$ & $C_{\rm c}$  & $I_{\rm c}(0)$ & EWSR \\
 & ($e^2$MeVfm$^4$) & (MeVfm$^5$) & ($e^2$/fm) & ($e^2$MeVfm$^4$)\\
\hline
SkM$^\ast$ & 15587 & 68.74 & 58.01 & 19575\\
SLy4 & 15722 & 32.47 & 57.90 & 17602\\
SkP & & 44.64 & & 
\end{tabular}
\end{ruledtabular}
\end{table}

\clearpage

\subsection{\label{subsec:strength function  1-}The 1$^{\bm -}$ channels}

\begin{figure} 
\includegraphics[width=1.0\textwidth]{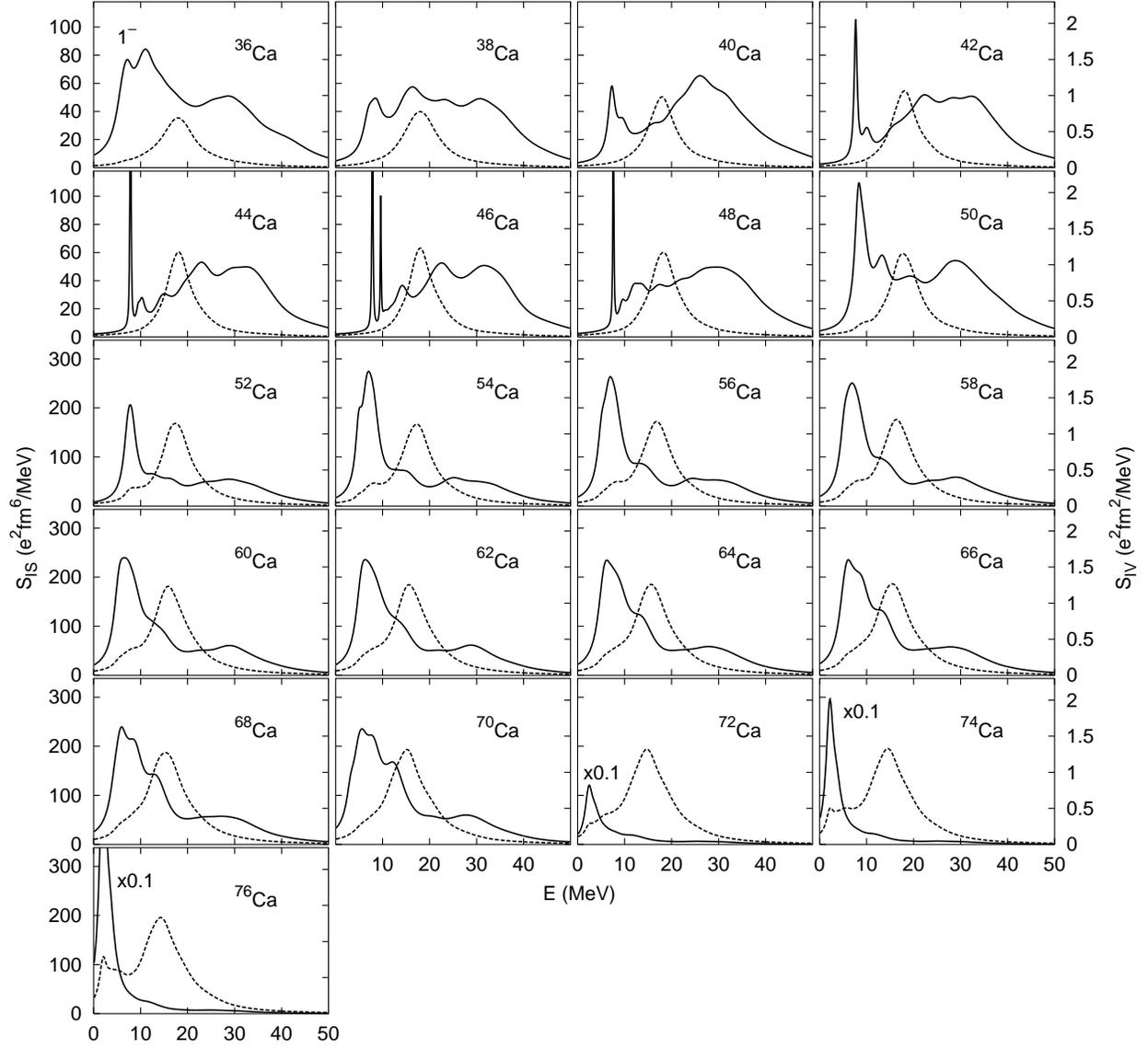}
\caption{\label{fig:str_ca_1-}
IS (solid, scale on left)
and IV (dashed, scale on right)
1$^-$ strength functions for even Ca isotopes (SkM$^\ast$).
The IS strength functions in $^{72,74,76}$Ca are reduced by a factor of 10.}
\end{figure}

Having analyzed the $0^+$ strength functions in detail, we
will, for the most part, let the higher-multipole figures speak for themselves.
Just a few remarks about the $1^-$ channel:  The IS strength functions in Ca
appear in Fig.~\ref{fig:str_ca_1-}.  Here, unlike in the $0^+$ channel, there
is a clear low-energy peak in all nuclei, increasing in
size\footnote{The low-energy peaks of $^{42-48}$Ca are so sharp because we
smooth the strength functions with a smaller width
at bound states than in the continuum (see Ref.~\cite{Ter05}).}
after $^{48}$Ca. (The IS strength
functions of $^{72,74,76}$Ca are multiplied by 0.1 in the figure.) In $^{76}$Ca
45 \% of the energy-weighted strength is below 10 MeV.
The IV strength functions also have increasing low-energy strength at large
$N$.  The small low-energy bump in $^{50}$Ca is produced by SLy4 as well as
SkM$^{\ast}$.

\begin{figure} 
\includegraphics[width=1.0\textwidth]{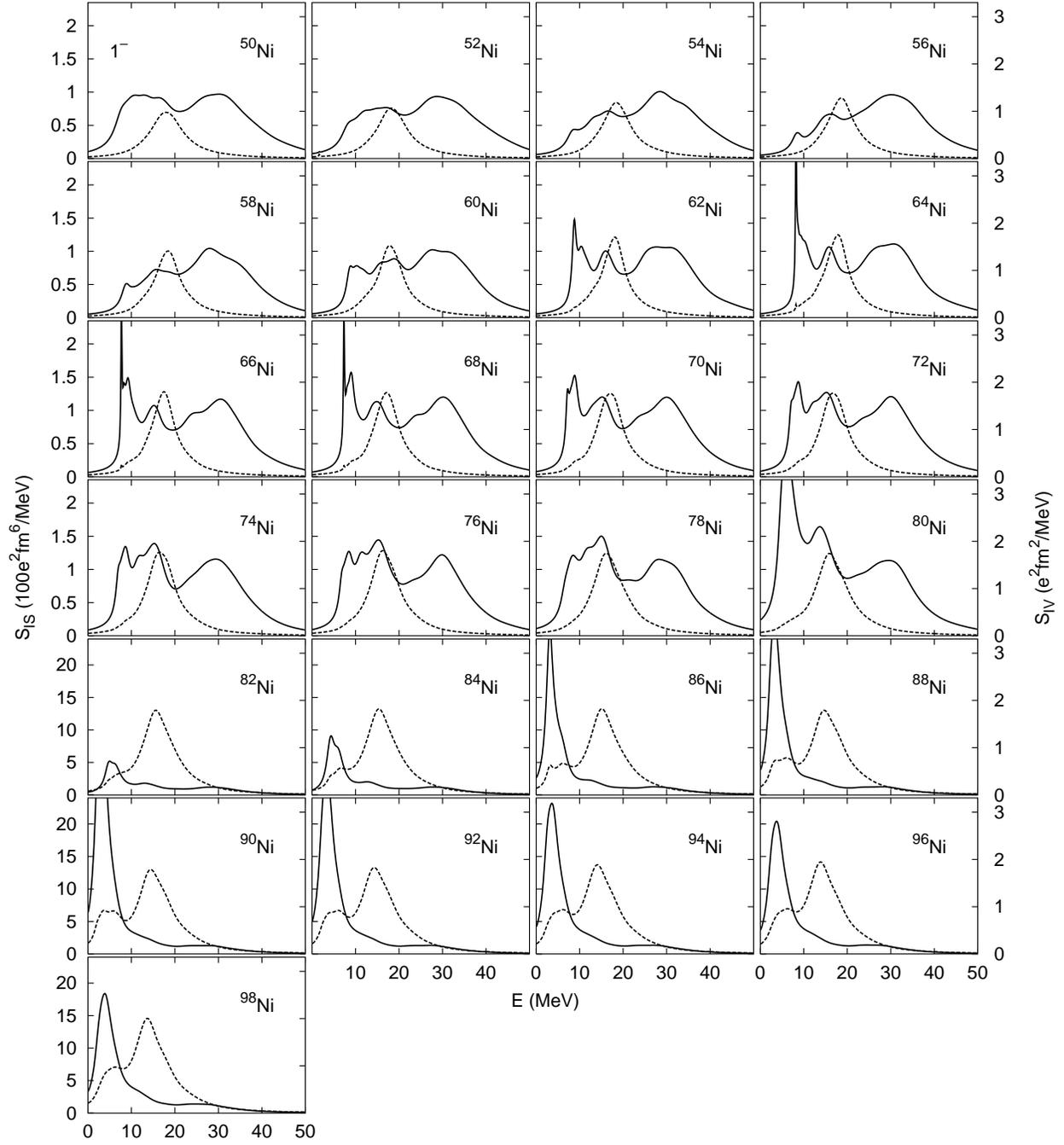}
\caption{\label{fig:str_ni_1-} 
IS (solid, scale on left)
and IV (dashed, scale on right)
1$^-$ strength functions for even Ni isotopes (SkM$^\ast$).
 }
\end{figure}

\begin{figure}
\includegraphics[width=1.0\textwidth]{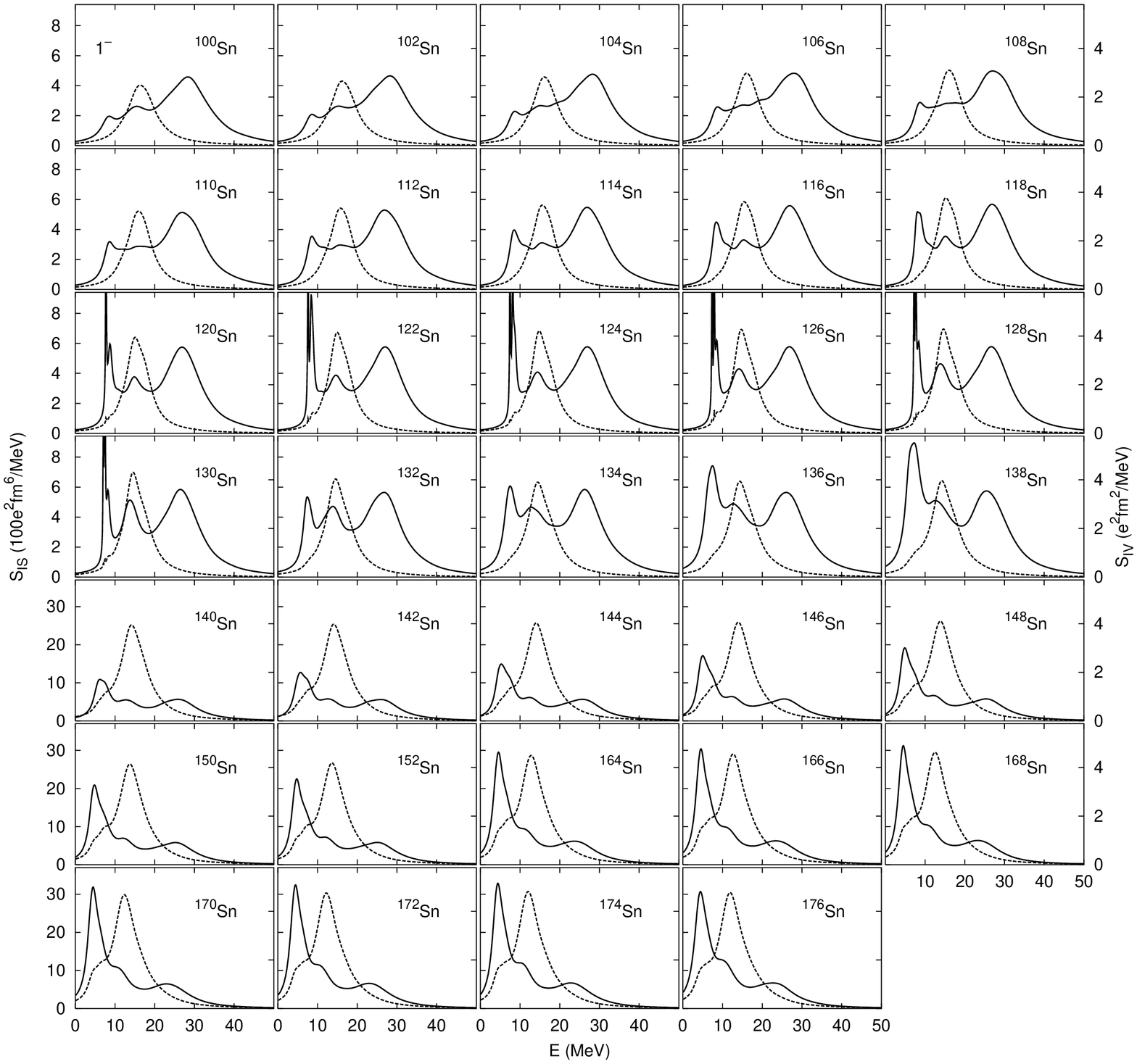}
\caption{\label{fig:str_sn_1-} 
The same as Fig.~\ref{fig:str_ni_1-} but for Sn isotopes.
 }
\end{figure}

We see similar $N$-dependence in Ni (see Fig.~\ref{fig:str_ni_1-}), though after
$^{90}$Ni the low-energy IS peak shrinks.  The threshold for rapid growth with
both Skyrme interactions is the closed-shell nucleus $^{78}$Ni.  In Sn,
once again, the low-energy IS peak grows with $N$, though more smoothly than in
the other isotopes (see Fig.~\ref{fig:str_sn_1-}).  In particular, there is
apparently no sudden increase at the closed-shell-plus-two nucleus $^{134}$Sn.
Recently, Ref.\ \cite{Adr05} reported a clear IV bump around 10 MeV in that
nucleus.  The
bumps in our calculated spectra do not seem to really set in until higher $N$.

\begin{figure}
\includegraphics[width=0.8\textwidth]{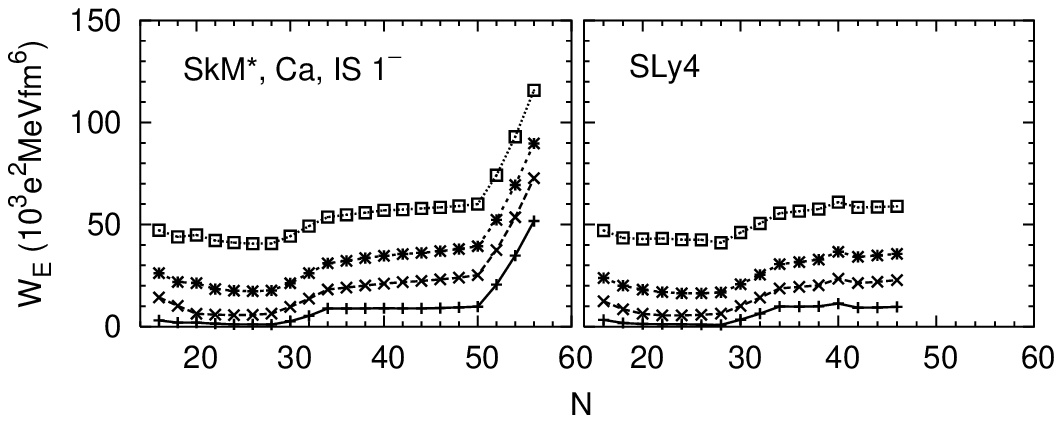}
\caption{\label{fig:part_esum_is_ca1-} 
The same as Fig.~\ref{fig:part_esum_is_ca0+} but for the $1^-$ channel.
 }
\end{figure}

\begin{figure}
\includegraphics[width=0.8\textwidth]{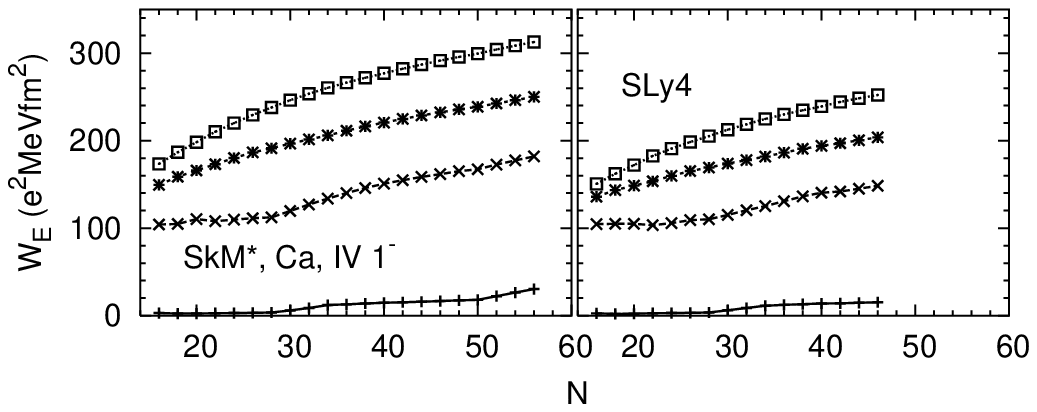}
\caption{\label{fig:part_esum_iv_ca1-} 
The same as Fig.~\ref{fig:part_esum_iv_ca0+} but for the $1^-$ channel.
 }
\end{figure}

\begin{figure}
\includegraphics[width=0.8\textwidth]{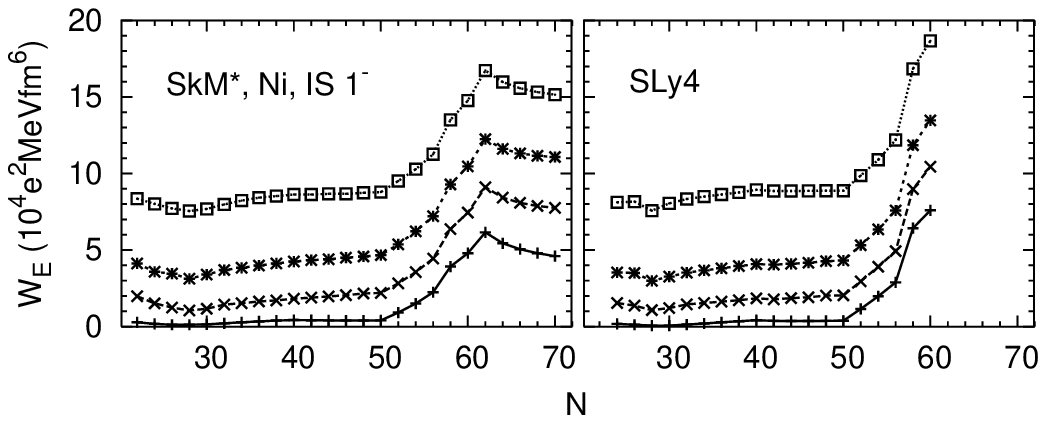}
\caption{\label{fig:part_esum_is_ni1-} 
The same as Fig.~\ref{fig:part_esum_is_ni0+} but for the $1^-$ channel.
 }
\end{figure}

\begin{figure}
\includegraphics[width=0.8\textwidth]{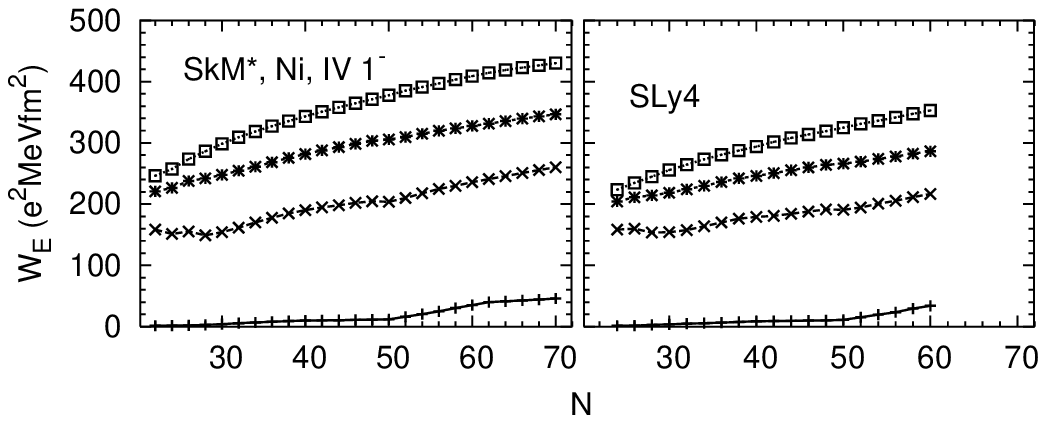}
\caption{\label{fig:part_esum_iv_ni1-} 
The same as Fig.~\ref{fig:part_esum_iv_ni0+} but for the $1^-$ channel.
 }
\end{figure}

\begin{figure}
\includegraphics[width=1.0\textwidth]{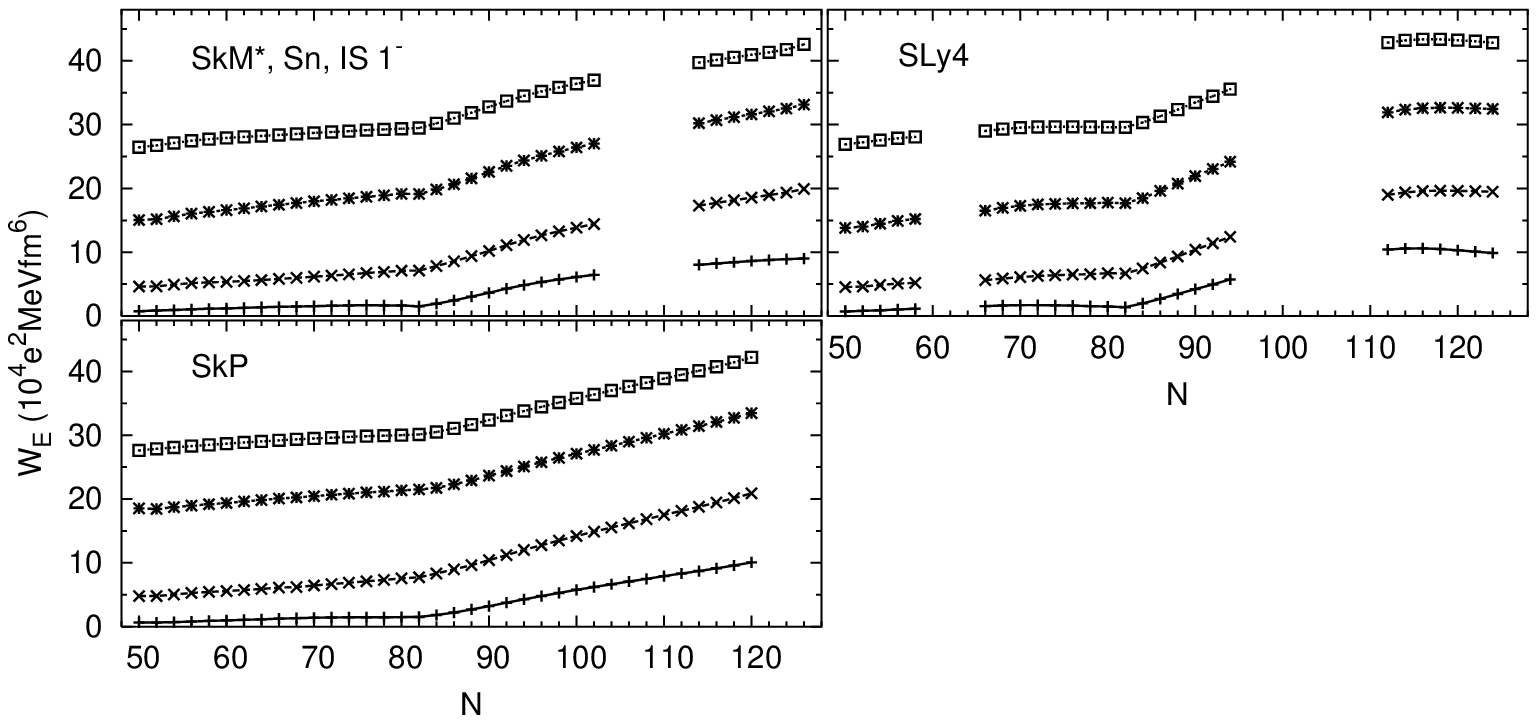}
\caption{\label{fig:part_esum_is_sn1-} 
The same as Fig.~\ref{fig:part_esum_is_sn0+} but for the $1^-$ channel.
 }
\end{figure}

\begin{figure}
\includegraphics[width=1.0\textwidth]{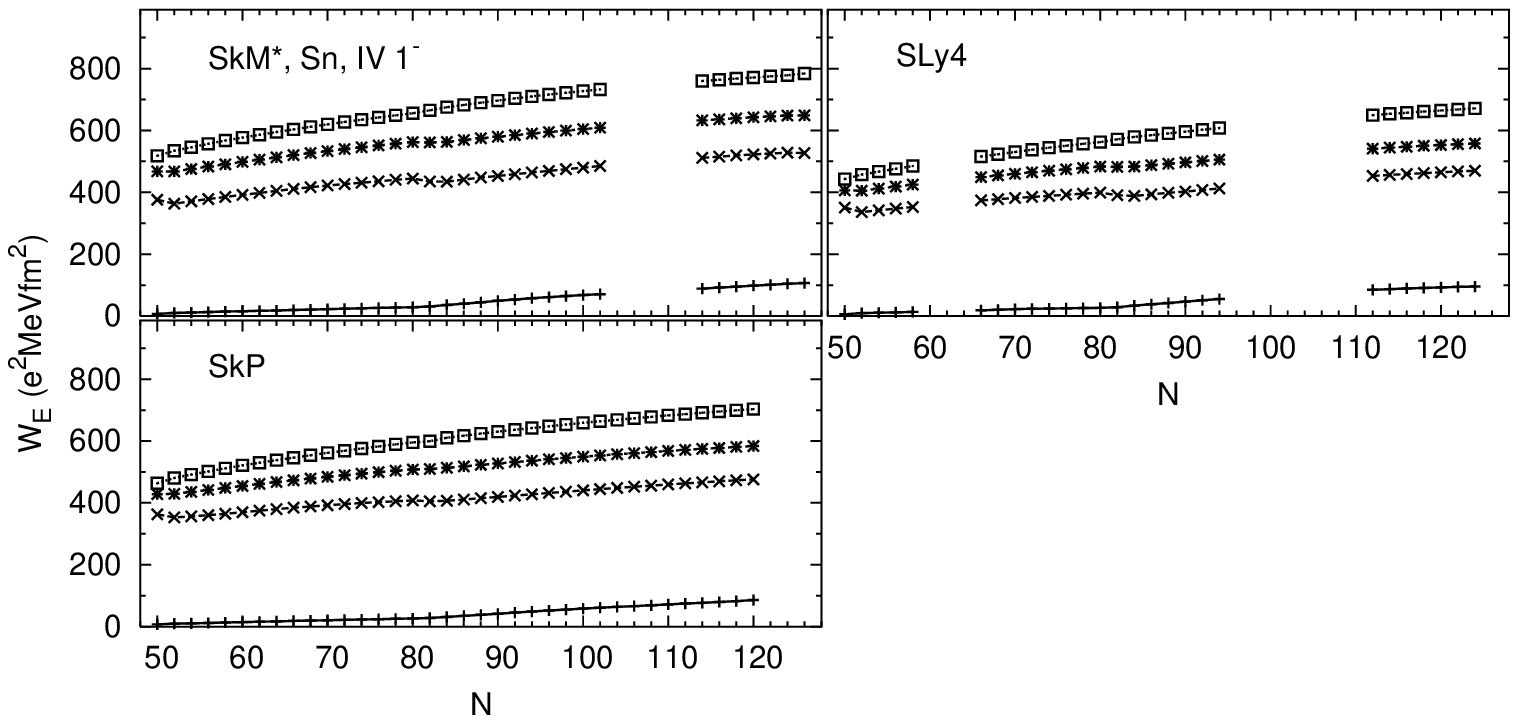}
\caption{\label{fig:part_esum_iv_sn1-} 
The same as Fig.~\ref{fig:part_esum_iv_sn0+} but for $1^-$ channel.
 }
\end{figure}

The $1^-$ $W_E$ tell a slightly different story
(Figs.~\ref{fig:part_esum_is_ca1-}--\ref{fig:part_esum_iv_sn1-}).  The partial
sums in the IS channel of Sn (Fig.~\ref{fig:part_esum_is_sn1-}) show a
noticeable kink at $N=82$, due clearly to an increase in low-energy strength.
As remarked above, the increase is not obvious in the strength-function
figures.  The choice of interaction makes very little difference in the $W_E$
except in the blank regions (indicating quadrupole deformation)  and near the
neutron drip line. 

According to the analytical argument in the previous subsection, if the IV EWSR
depends on the interaction in one multipole, then it should in all
multipoles.  As Figs.~\ref{fig:part_esum_iv_ca1-}, \ref{fig:part_esum_iv_ni1-},
and \ref{fig:part_esum_iv_sn1-} show, this is indeed  the case. 

We have not commented on the question of whether protons are involved in the
transitions to low-lying states in this channel because the IS and IV operators
do not have the same radial dependence.  We will address the issue in later when
we display transition densities.

\clearpage


\clearpage

\subsection{\label{subsec:strength function 2+}The 2$^{\bm +}$ channels}

\begin{figure}
\includegraphics[width=1.0\textwidth]{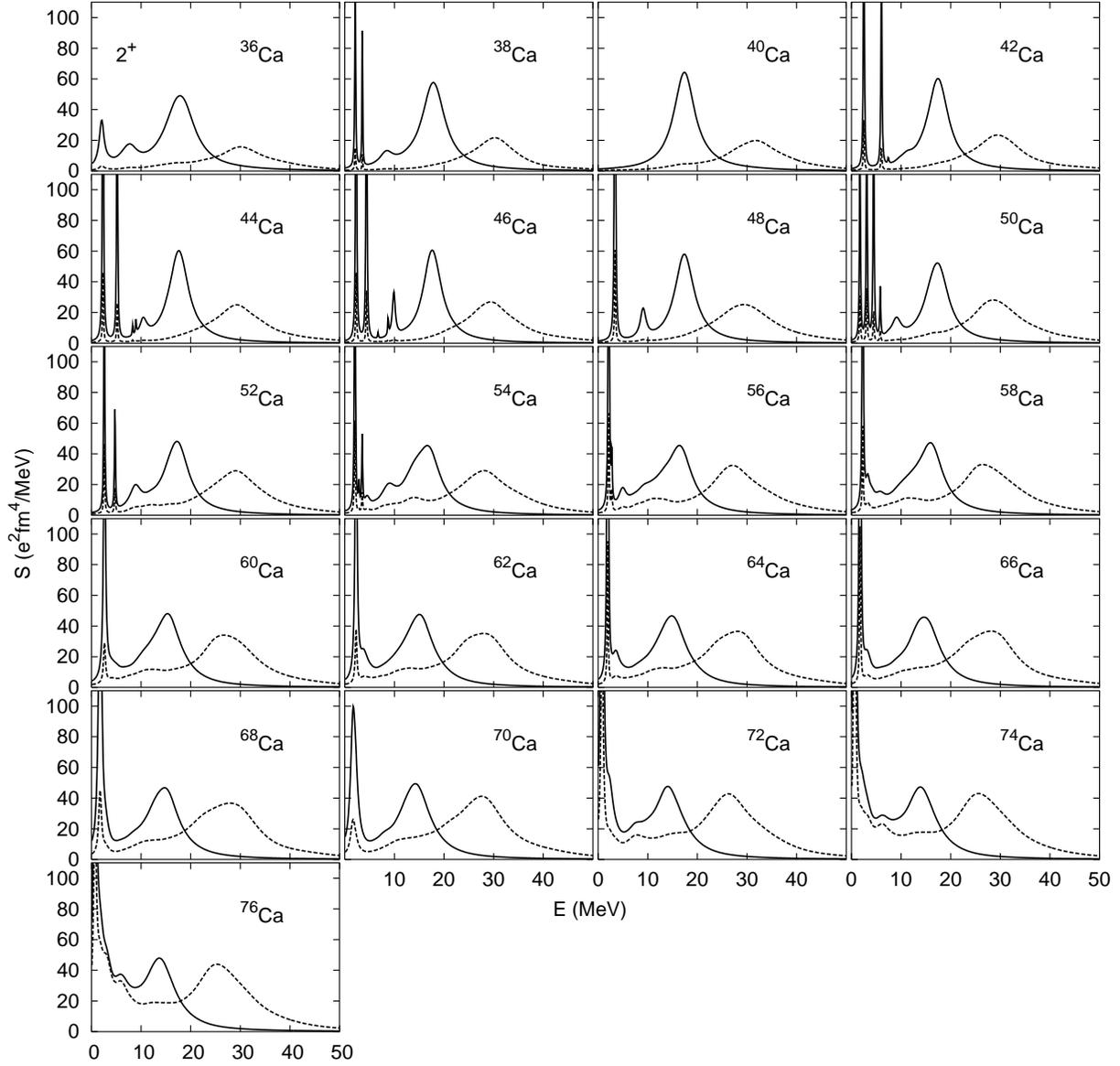}
\caption{\label{fig:str_ca_2+} 
IS (solid) and IV (dashed)
2$^+$ strength functions in even Ca isotopes (SkM$^\ast$).
 }
\end{figure}

The strength functions in the $2^+$ channels are qualitatively different from
those in
the lower-multipole channels.  In Ca, for example 
(Fig.~\ref{fig:str_ca_2+}), all isotopes\footnote{Reference \cite{Sag01}
calculates the $2^+$ strength functions of $^{60}$Ca
without low-energy peaks; we do not know the reasons for the difference between
that calculation and ours.}
except $N=40$ have considerable
low-energy strength, in many instances because of sharp quadrupole vibrations.
The low-energy strength grows with $N$, but unlike in the $0^+$ channel, both neutrons
and protons are involved in the transitions to these states, even near the drip
line, because IS and IV distributions do not coincide.
This is so for all Skyrme interactions.  In fact the IS strength at low energies
is much larger than the IV strength in all isotopes except $^{76}$Ca.
Even there,
the IS strength to the low-energy peak
with SkM$^{\ast}$ is 256 $e^2$fm$^4$ and the IV strength only 136 $e^2$fm$^4$.

\begin{figure}
\includegraphics[width=1.0\textwidth]{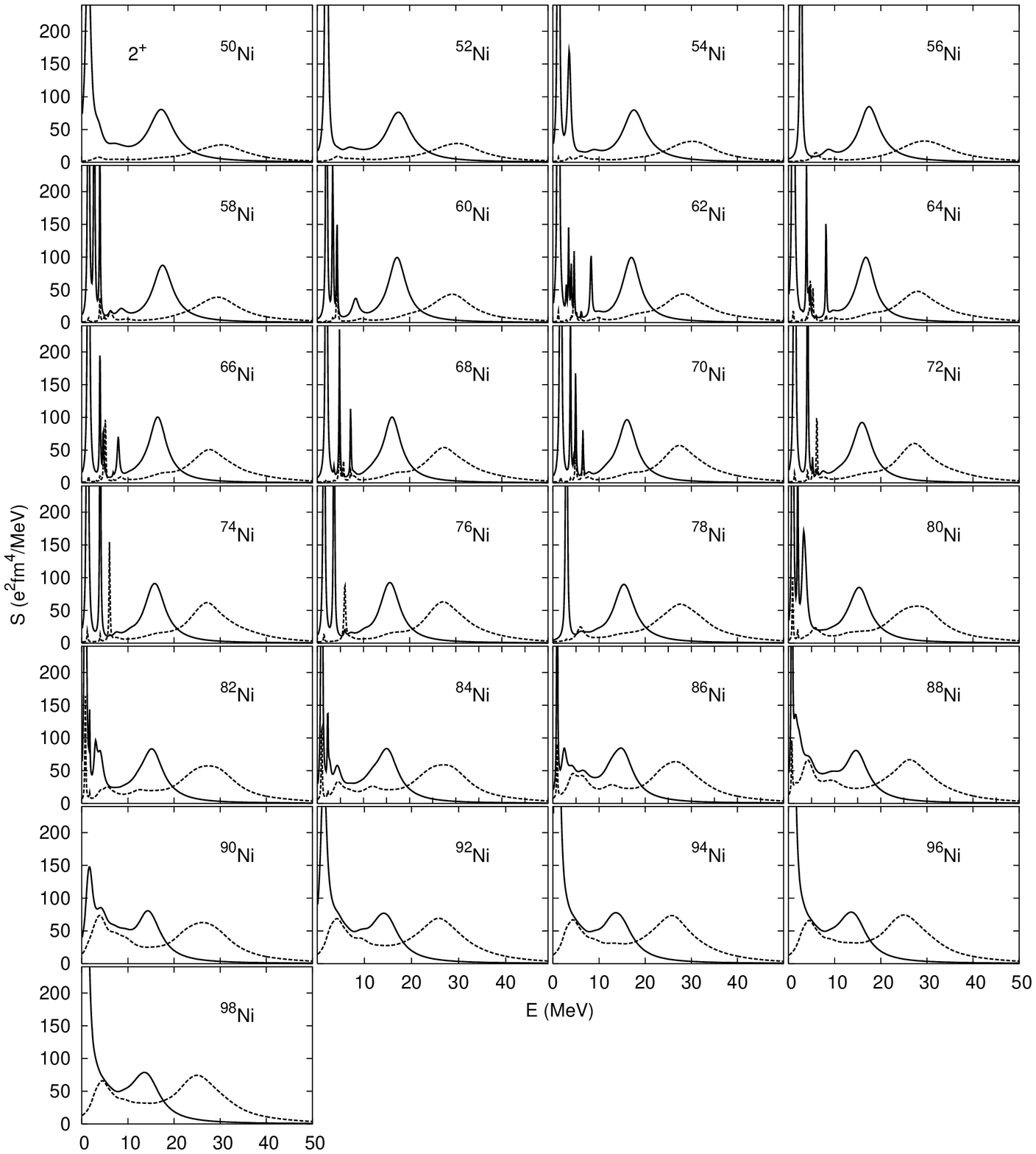}
\caption{\label{fig:str_ni_2+} 
The same as Fig.~\ref{fig:str_ca_2+} but for Ni isotopes.
 }
\end{figure}

The  $2^+$ strength functions in Ni (Fig.~\ref{fig:str_ni_2+}) 
are similar, though with two differences:  
all the nickel isotopes have low-energy peaks, and for $^{90-98}$Ni  
the IV peak is at a different energy than the IS peak.  The lowest-lying
peak seems to be nearly pure IS, like the familiar surface-quadrupole vibrations
in stable nuclei. In the next section we will discuss the extent to which the
states near the neutron-drip line
display the same vibrational features we observe near stability.

SLy4 puts the
drip line at $^{88}$Ni, much closer to stability, and in isotopes close to 
$^{88}$Ni  puts the
IS and IV peaks at about the same location.  In the lighter
($A \geq 80$) Ni isotopes, the lowest
peak always has a little bit of IV strength mixed with the IS 
strength, as with SkM$^\ast$. In
$^{80}$Ni (SkM$^\ast$), the IV strength in the peak at 400 keV is 5 \% of the
IS strength.

\begin{figure}
\includegraphics[width=1.0\textwidth]{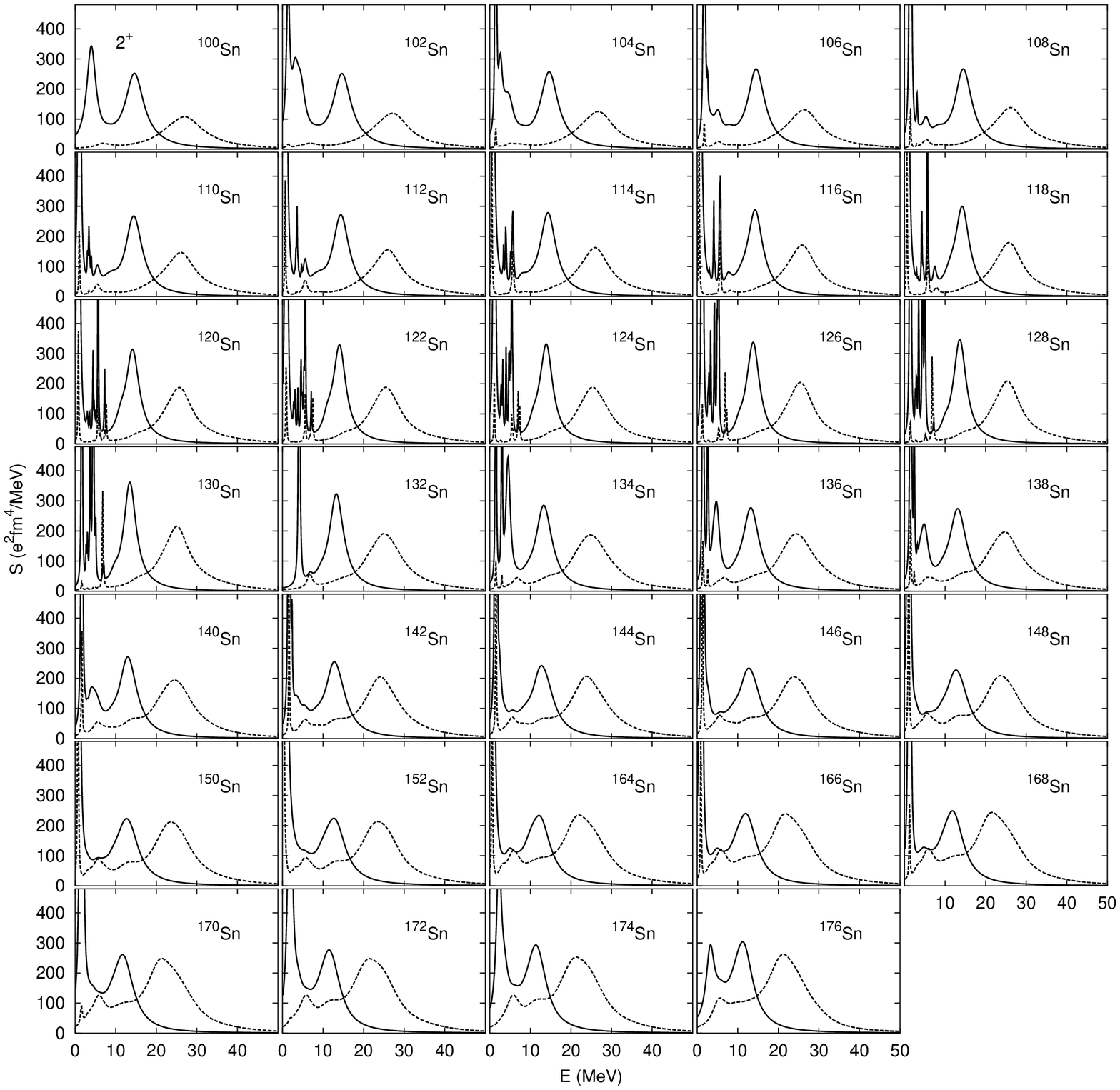}
\caption{\label{fig:str_sn_2+} 
The same as Fig.~\ref{fig:str_ca_2+} but for Sn isotopes.
 }
\end{figure}

Things are quite similar in Sn (Fig.~\ref{fig:str_sn_2+}); for $^{134-170}$Sn
the lowest-energy peak has both IS and the IV components but in $^{172-176}$Sn
it has no IV component.  The other interactions give the same results near the
neutron drip line, indicating again that proton excitations are important in
these modes everywhere.  

\begin{figure}
\includegraphics[width=0.8\textwidth]{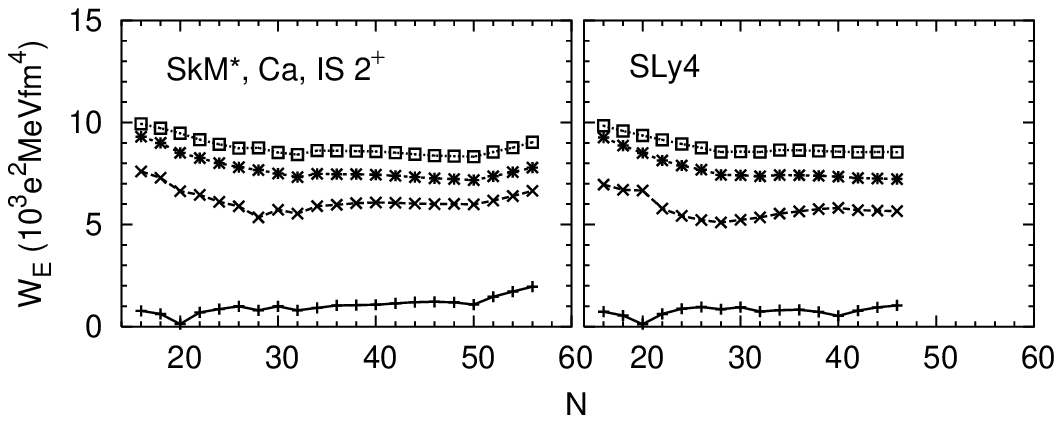}
\caption{\label{fig:part_esum_is_ca2+} 
The same as Fig.~\ref{fig:part_esum_is_ca0+} but for the $2^+$ channel.
 }
\end{figure}

\begin{figure}
\includegraphics[width=0.8\textwidth]{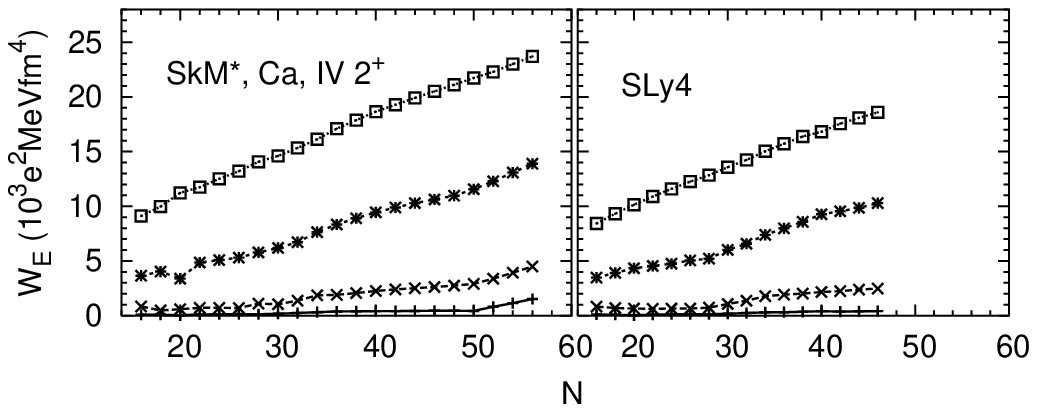}
\caption{\label{fig:part_esum_iv_ca2+} 
The same as Fig.~\ref{fig:part_esum_iv_ca0+} but for the $2^+$ channel.
 }
\end{figure}

The predictions for $W_E$ in Ca (Fig.~\ref{fig:part_esum_is_ca2+}) show that
SLy4 evinces slightly more of a closed shell (a
small kink in the 10-MeV curve) at $N=40$ than does SkM$^{\ast}$, while showing
less of a shell effect at $N = 28$ (no small kink in the 20-MeV curve).  As was
the case in the $0^+$ and $1^-$ channels, the IV $W_E$
(Fig.~\ref{fig:part_esum_iv_ca2+}) depends on the interaction.

\begin{figure}
\includegraphics[width=0.8\textwidth]{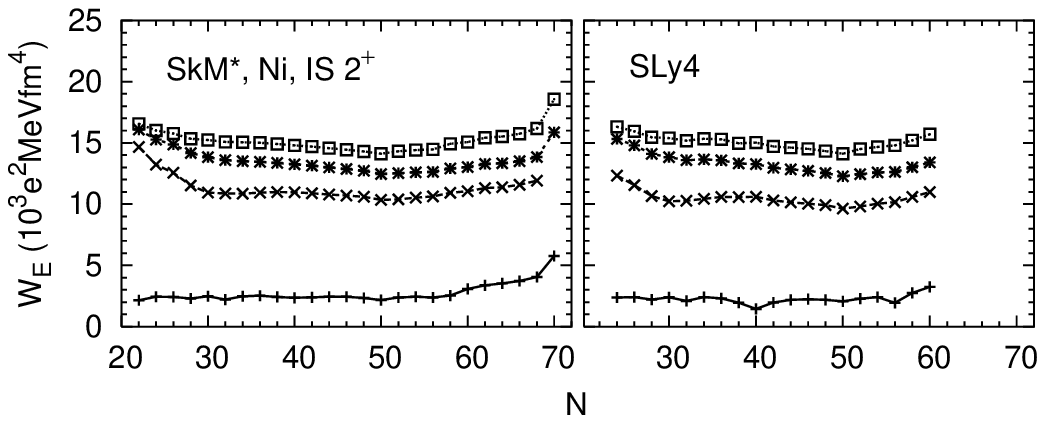}
\caption{\label{fig:part_esum_is_ni2+} 
The same as Fig.~\ref{fig:part_esum_is_ni0+} but for the $2^+$ channel.
 }
\end{figure}

\begin{figure}
\includegraphics[width=0.8\textwidth]{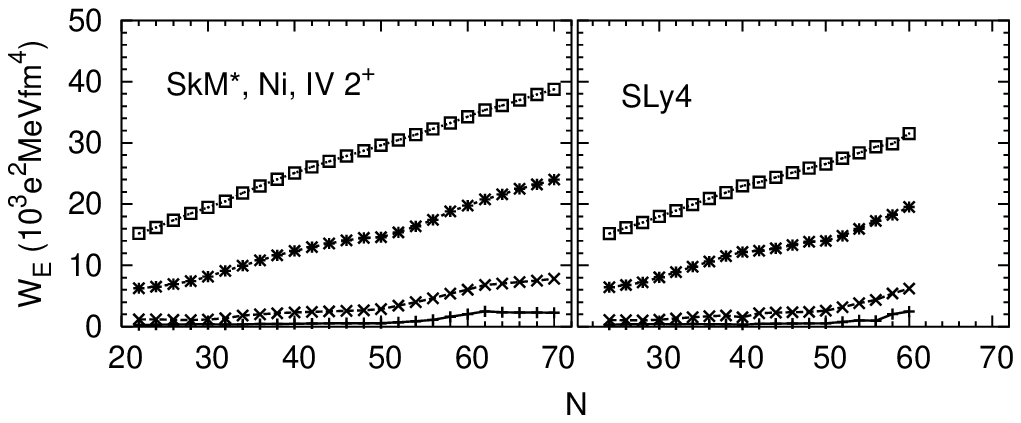}
\caption{\label{fig:part_esum_iv_ni2+} 
The same as Fig.~\ref{fig:part_esum_iv_ni0+} but for the $2^+$ channel.
 }
\end{figure}

\begin{figure}
\includegraphics[width=1.0\textwidth]{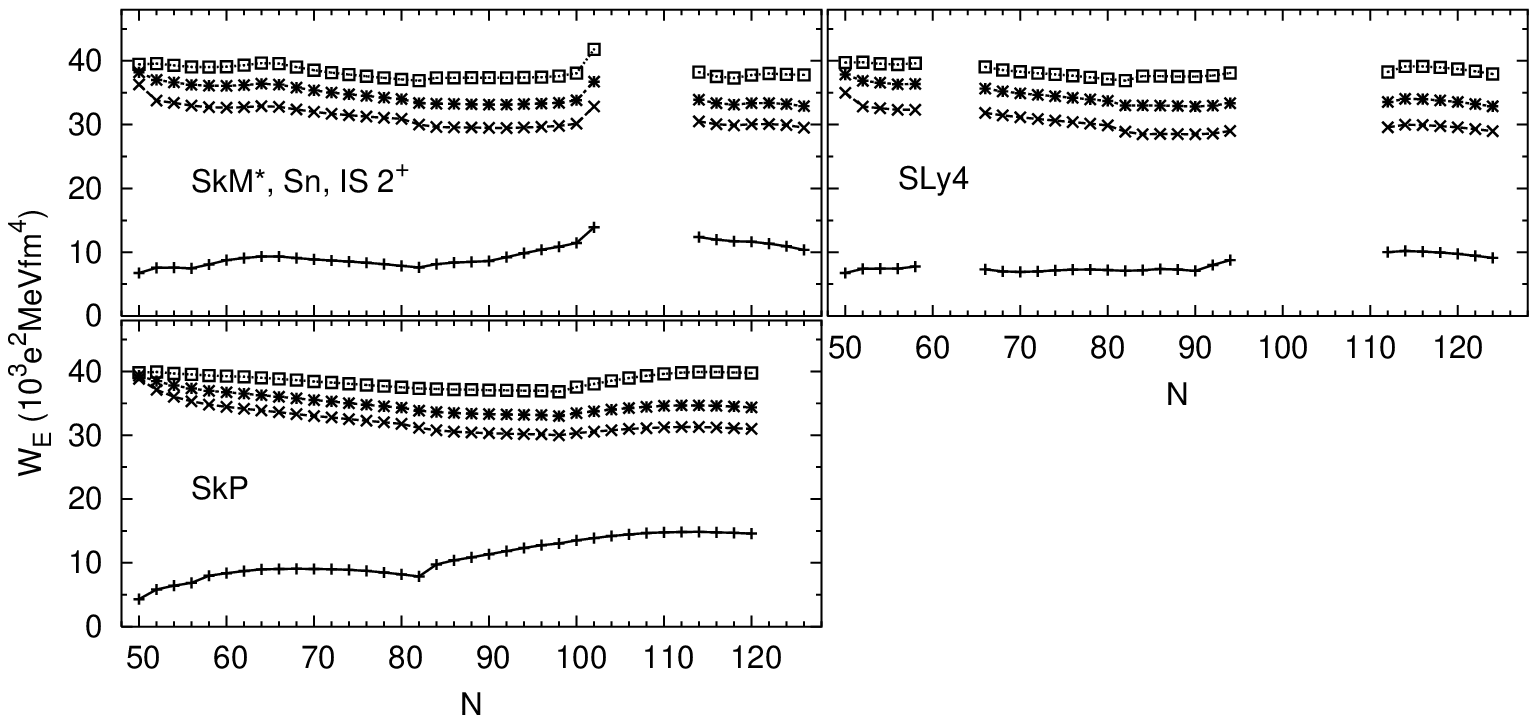}
\caption{\label{fig:part_esum_is_sn2+} 
The same as Fig.~\ref{fig:part_esum_is_sn0+} but for the $2^+$ channel.
 }
\end{figure}

\begin{figure}
\includegraphics[width=1.0\textwidth]{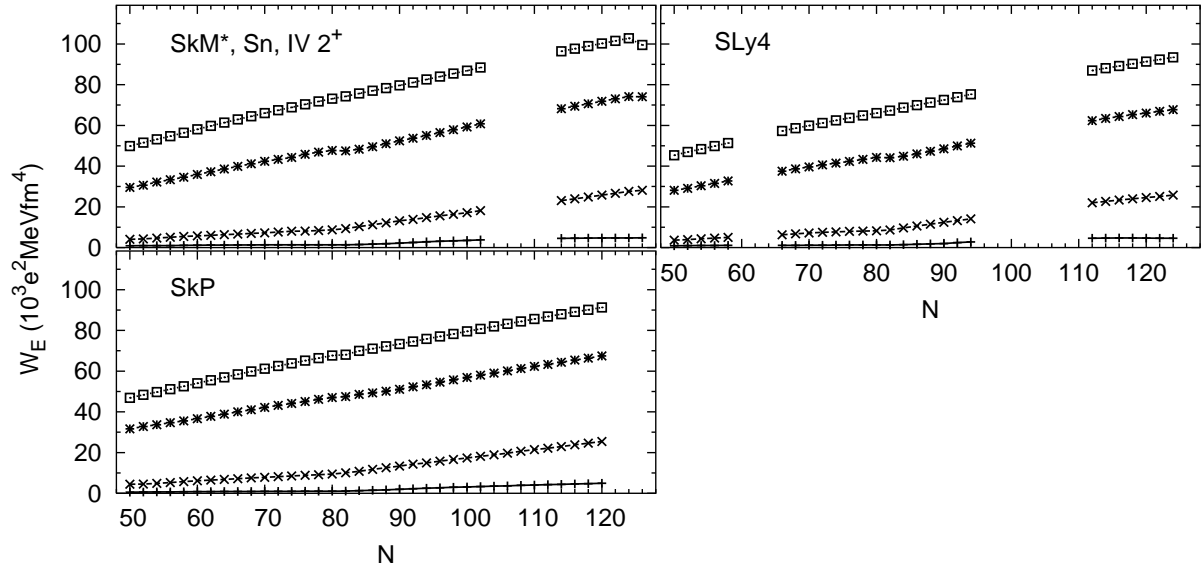}
\caption{\label{fig:part_esum_iv_sn2+} 
The same as Fig.~\ref{fig:part_esum_iv_sn0+} but for $2^+$ channel.
 }
\end{figure}

Similar differences in shell-strength among the Skyrme interactions are
apparent in the Ni and Sn $W_E$ curves for the IS channel
(Figs.\ \ref{fig:part_esum_is_ni2+} and \ref{fig:part_esum_is_sn2+}), 
while the difference is not obvious in the IV channels
(Figs.~\ref{fig:part_esum_iv_ni2+} and \ref{fig:part_esum_iv_sn2+}). 
The IS $W_E$ for Sn
(Fig.~\ref{fig:part_esum_is_sn2+}) indicate that SkP has more low-energy
strength $(N>82)$ than the other parameter sets. The IV curves in Sn show
interaction differences similar to those in the $1^-$ channel. 

\clearpage

\begin{figure}
\end{figure}

\clearpage

\section{\label{sec:transition density}Transition densities} 
Giant resonances lie in the continuum and their
$N$-dependence is therefore smooth.  The low-energy peaks, on the other hand, 
can vary dramatically with $N$, especially near the drip line.  In this
section, to learn more
about their collectivity and $N$ dependence, we examine transition densities to
representative low-lying states.
	
The  transition density 
$\rho_{\rm tr}^q(\bm{r};k)$ is defined in the Appendix. 
Here we focus on the radial transition density
\begin{equation}
\rho_{\rm tr}^q(r;k) = r^2\int d\Omega\,Y_{J_kM_k}(\Omega)
\rho_{\rm tr}^q(\bm{r};k).
\end{equation}
The transition amplitude depends on this quantity through the relation
\begin{equation}
\langle k|\hat{F}_{JM}|0\rangle = 
\int dr\,\sum_{q={\rm p}, {\rm n}}  f^q(r)\rho_{\rm tr}^q(r;k), 
\end{equation}
where $f^q(r)$ is the radial part of the multipole operator
$\hat{F}_{JM}$.  The radial transition density therefore specifies the radius
at which the strength is large or small.  The transition density, as
Eq.~(\ref{eq:tr_den}) shows, is proportional to the product of particle
and hole wave functions and so is localized as long as all hole states are localized,
i.e.\ as long as the ground state is bound.
In this section, as before, we will be
using SkM$^\ast$ unless we state otherwise.

\subsection{\label{subsec:transition density 0+}The 0$^{\bm +}$ channels}

\begin{figure}
\includegraphics{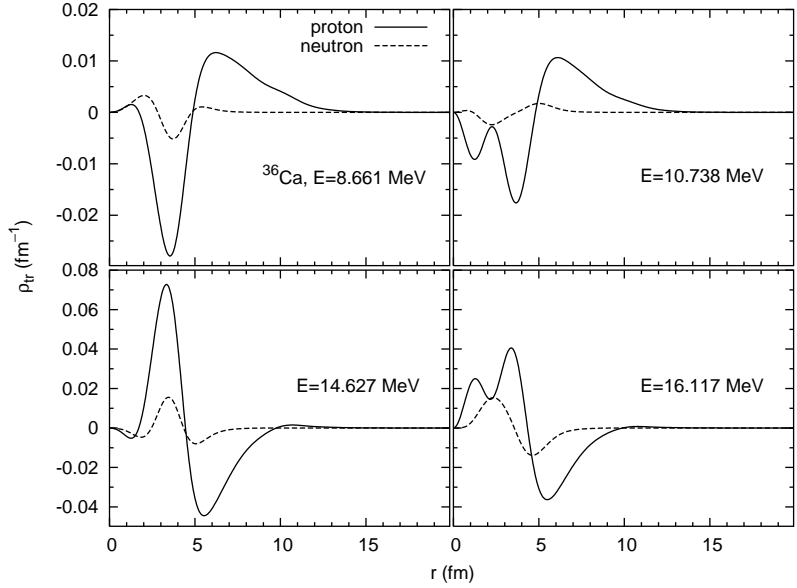}
\caption{\label{fig:tr_den_ca0+n16} 
Proton and neutron transition densities to the four states in $^{36}$Ca 
in the peaks of the strength function ($E<17$ MeV) in
Fig.~\ref{fig:str_ca_0+}.
 }
\end{figure}

Figure \ref{fig:tr_den_ca0+n16} shows the transition densities to 
4 states below the giant resonances in $^{36}$Ca with large strengths (the energies label the
results of the QRPA calculation before smoothing).
Proton excitations are important, particularly because of the factor $r^2$
in the $0^+$  transition operator.
The rms radii of the protons and neutrons in the ground state are 3.40 and 3.19 fm.
The important part for the strength comes from outside
the rms radius.  

\begin{figure}
\includegraphics{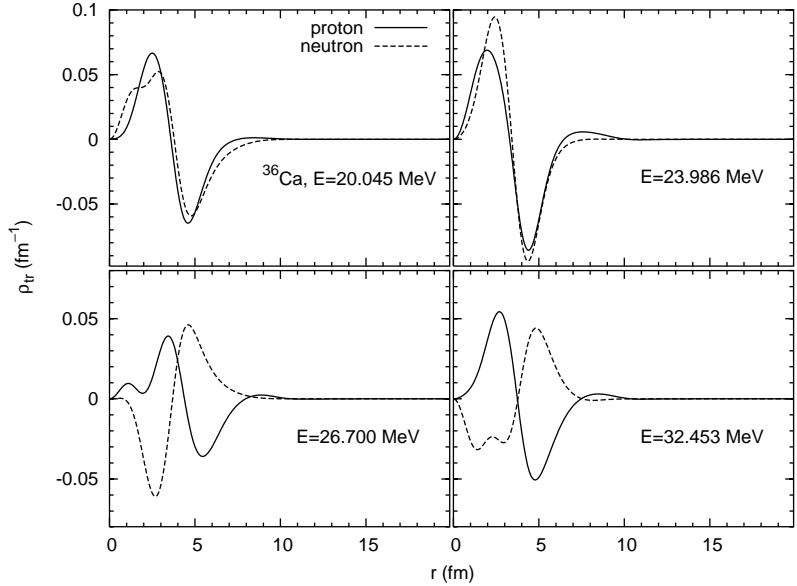}
\caption{\label{fig:tr_den_gr_ca0+n16} 
The same as Fig.~\ref{fig:tr_den_ca0+n16} but for 
the giant resonances. The states in the upper panels 
correspond to the peaks of the IS giant resonance, and in
the lower to the IV giant resonance.
 }
\end{figure}

This is not true of the giant resonances.
Figure \ref{fig:tr_den_gr_ca0+n16} shows transition densities in
the peaks of the giant resonances of $^{36}$Ca. 
The upper two panels represent the IS giant resonance,
and as is expected the protons and neutrons have similar transition densities. 
In the lower panels, corresponding to the IV giant resonance, 
the two densities are out of phase.  
The peaks of the proton transition densities are closer to the center of the
nucleus than those of the
lower-energy 
states, particularly in the IS channel. 

\begin{figure}
\includegraphics{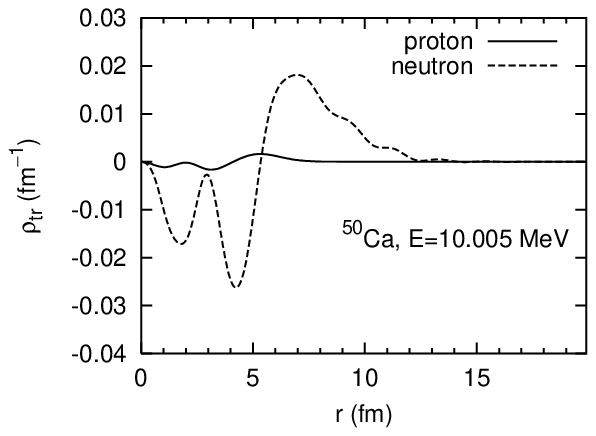}
\caption{\label{fig:tr_den_ca0+n30} 
Transition densities to the excited state in 
the small low-energy bump in the $0^+$ strength function of $^{50}$Ca.
 }
\end{figure}

We saw that a small bump appeared after magic numbers in neutron-rich nuclei.
In Ca, the relevant isotope is $^{50}$Ca.  
(Fig.~\ref{fig:tr_den_ca0+n30}) shows
that the excitation is created entirely by neutrons.
But the state is not at all collective.
The largest two-quasiparticle component in the QRPA wave function is
$\nu3p_{3/2}$(${\cal E} = 9.175$ MeV, 
$\bar{v}_\mu^2 = 6.0\times 10^{-4}) 
 \otimes \nu2p_{3/2}$(${\cal E} = 0.880$ MeV, $\bar{v}_\mu^2 =
0.488)$, where ${\cal E}$ is the 
quasiparticle energy in the HFB quasiparticle basis
and $\bar{v}_\mu^2$ is the diagonal ground-state occupation probability 
(norm of the lower component of the 
quasiparticle wave function).  
The radial quantum numbers are assigned in order of 
${\cal E}$ ($-{\cal E}$ for those having $\bar{v}_\mu^2 > 0.5$). 
$\bar{X}_{\mu\nu}$($\bar{Y}_{\mu\nu}$) is the forward (backward) amplitude of 
the QRPA solution.

The squared amplitude
$\bar{X}_{\mu\nu}^2-\bar{Y}_{\mu\nu}^2$
of the $\nu3p_{3/2}\otimes\nu2p_{3/2}$ component is $0.980$.
The component corresponds to the excitation of a neutron from a
$p_{3/2}$ orbit near the Fermi surface to 
another $p_{3/2}$ orbit about 9 MeV above. 
The first orbit is strongly affected by pairing correlations.
The two-quasiparticle energy is 10.055 MeV, and the energy 
of the excited state itself is 10.005 MeV; the excitation is clearly almost
unaffected
by the residual interaction.  Yet this nearly pure two-quasiparticle state,
because the neutron transition density has such a long tail,
carries 5 \% of the total IS 
transition strength.  Furthermore, protons contribute essentially nothing to
the transition; the maximum of the $\bar{X}_{\mu\nu}^2-\bar{Y}_{\mu\nu}^2$
for protons is $0.3\times 10^{-4}$.  This remarkable situation  --- dominance
by a single  two-neutron-quasiparticle configuration with very large strength,
occurs in the $0^+$ channel repeatedly.  

\begin{figure}
\includegraphics{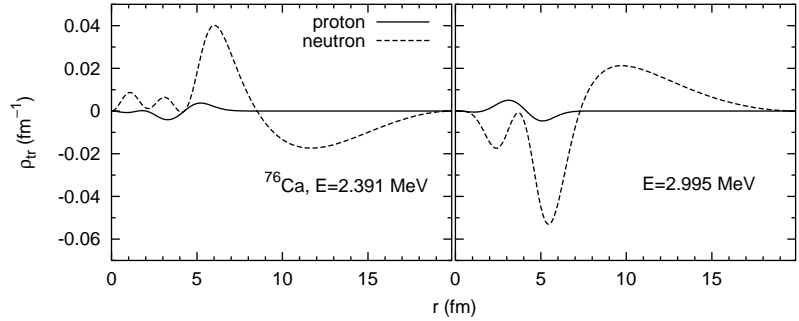}
\caption{\label{fig:tr_den_ca0+n56} 
Transition densities to the two excited states in
the low-energy peak in the $0^+$ strength functions of $^{76}$Ca.
 }
\end{figure}

Take, for example, the drip-line nucleus $^{76}$Ca.
Two low-lying states, at 
$E=2.391$ and 2.995 MeV, together account for 25.8 \% of 
the total IS strength  and 10.2 \% of the EWSR.
Transition densities, shown in
Fig.~\ref{fig:tr_den_ca0+n56}, are again dominated by neutrons, with outer
peaks at 9--11 fm.  
The rms ground-state radii are 3.738 fm for protons and 4.615 fm 
for neutrons, so the highest (inner) 
peaks of $\rho_{\rm tr}(r;k)$ are close to the neutron rms radius.
The main component of the excited state with
$E=2.391$ MeV, with a squared amplitude of 0.807, is $\nu 3s_{1/2}
({\cal E}=0.355$ MeV, $\bar{v}_\mu^2=0.645) \otimes \nu 4s_{1/2}
({\cal E}=2.092$ MeV, $\bar{v}_\mu^2=0.012$).
The main component of the state at $E=2.995$ MeV, with a squared amplitude of 
0.930, is 
$\nu 2d_{5/2}({\cal E}=0.617$ MeV, $\bar{v}_\mu^2 = 0.78)
\otimes 
\nu 3d_{5/2}({\cal E}=2.457$ MeV, $\bar{v}_\mu^2 = 6.1\times10^{-3})$. 
Like before, these are nearly pure two-neutron-quasiparticle states with long
tails in their product.

\begin{figure}
\includegraphics{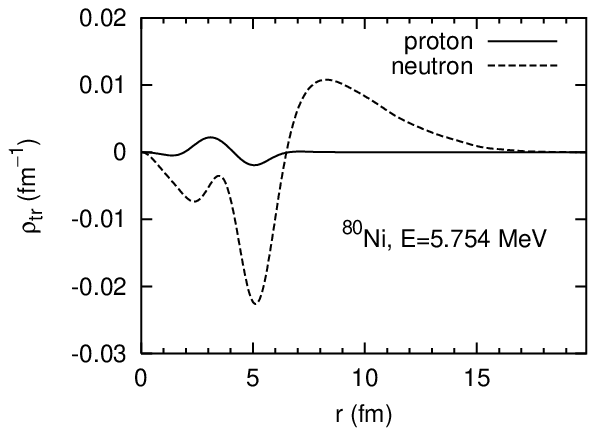}
\caption{\label{fig:tr_den_ni0+n52} 
Transition densities to the excited state in
the small low-energy bump in the $0^+$ strength function of $^{80}$Ni.
 }
\end{figure}

Turning now to Ni near the closed $N=50$ shell, we see that
$^{80}$Ni has a small bump around 5 MeV, while $^{78}$Ni 
does not (Fig.~\ref{fig:str_ni_0+}).
The transition density 
in Fig.~\ref{fig:tr_den_ni0+n52} again shows neutron dominance.
This excited state has 2.8 \% of the total IS strength, and its main component  is $\nu 2d_{5/2}({\cal E}=0.738$ MeV,
$\bar{v}_\mu^2=0.31)\otimes
\nu 3d_{5/2}({\cal E}=5.088$ MeV, $\bar{v}_\mu^2 = 7.1\times 10^{-4})$,
with $\bar{X}_{\mu\nu}^2-\bar{Y}_{\mu\nu}^2 = 0.761$. 
At the drip-line nucleus $^{98}$Ni, the main component of the excited state at
$E=3.452$ MeV, which is one 
of the states forming a low-energy bump, is 
$\{\nu 3d_{3/2}({\cal E}=1.799$ MeV, $\bar{v}_\mu^2=0.359)\}^2$; it has
$\bar{X}_{\mu\nu}^2-\bar{Y}_{\mu\nu}^2=0.818$. 

\begin{figure}
\includegraphics{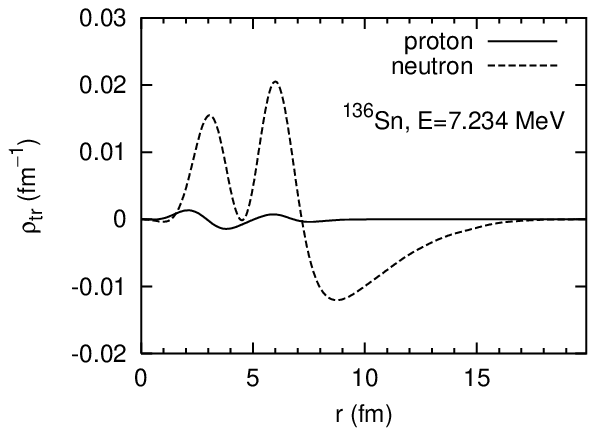}
\caption{\label{fig:tr_den_sn0+n86} 
Transition densities to the excited state in
a small bump in the $0^+$ strength function of $^{136}$Sn.
 }
\end{figure}

\begin{figure}
\includegraphics{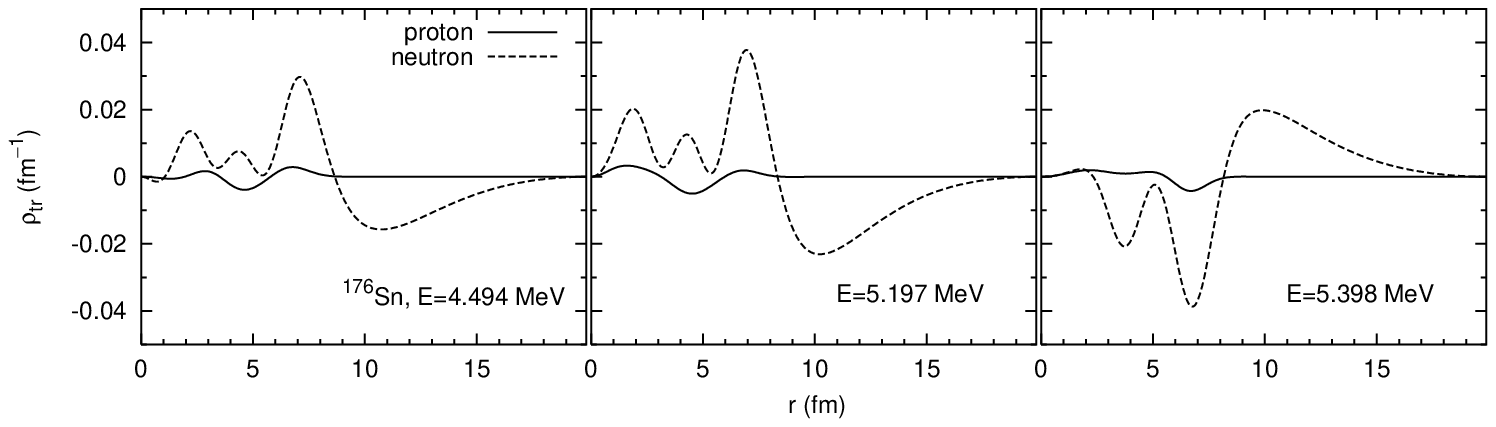}
\caption{\label{fig:tr_den_sn0+n126} 
Transition densities to the excited state in
a small bump in the $0^+$ strength function of $^{176}$Sn.
 }
\end{figure}

In Sn the nucleus just past the closed neutron shell is  $^{134}$Sn.  Here, as
we noted earlier, a new bump is not visible
(Fig.~\ref{fig:str_sn_0+}), until $^{136}$Sn.  We show the transition density
to the corresponding state in Fig.~\ref{fig:tr_den_sn0+n86}.  The excitation's
main component is $\nu 2f_{7/2}({\cal E}=0.950$ MeV, $\bar{v}_\mu^2=0.44)
\otimes \nu 3f_{7/2}({\cal E}=6.302$ MeV, $\bar{v}_\mu^2=6.2\times
10^{-4})$, with $\bar{X}_{\mu\nu}^2-\bar{Y}_{\mu\nu}^2=0.902$. 
${\cal E}=0.950$ MeV is
the smallest neutron quasiparticle energy of the HFB calculation.  Figure 
~\ref{fig:tr_den_sn0+n126} shows the transition densities to three states in
the low-energy bump in $^{176}$Sn.  The behavior is similar to that in the
drip-line Ca and Ni nuclei.

\begin{figure}
\includegraphics{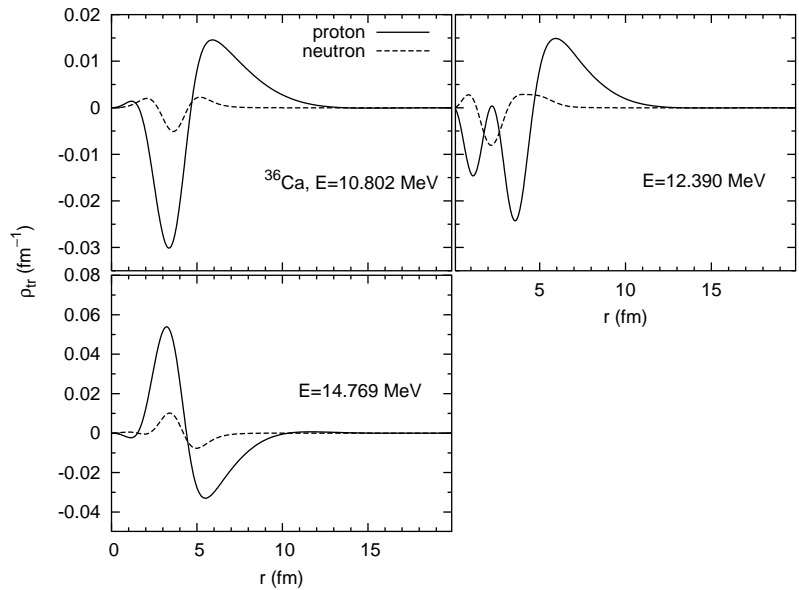}
\caption{\label{fig:tr_den_ca0+n16_sly4_skoprime} 
Transition densities to the $0^+$ excited states below the giant resonances in $^{36}$Ca with 
relatively large strength (SLy4).
}
\end{figure}

\begin{figure}
\includegraphics{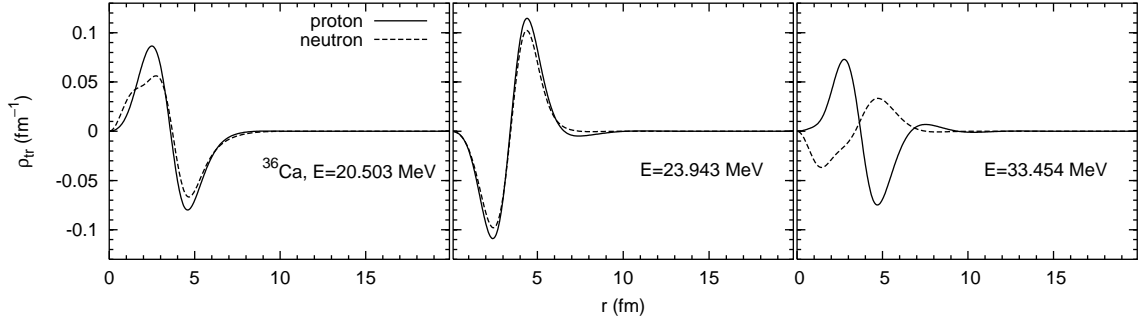}
\caption{\label{fig:tr_den_ca0+n16_gr_sly4_skoprime} 
Transition densities to the $0^+$ excited states of $^{36}$Ca in the
giant resonances (SLy4). 
}
\end{figure}

We turn finally to the dependence of the transition densities on the Skyrme
interaction.  Figure \ref{fig:tr_den_ca0+n16_sly4_skoprime} shows,
unsurprisingly by now, that SkM$^\ast$ and SLy4
give similar
transition densities to states below the giant resonances in $^{36}$Ca.
We show transition densities for some giant-resonance states in
Fig.~\ref{fig:tr_den_ca0+n16_gr_sly4_skoprime}.
The IS/IV character of those states is clear.

\begin{figure}
\includegraphics{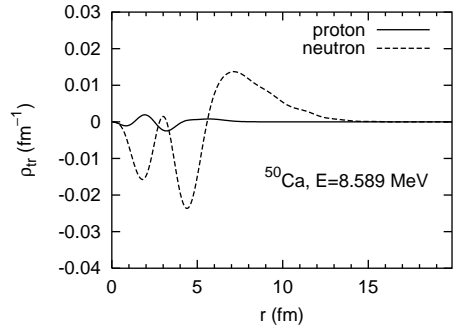}
\caption{\label{fig:tr_den_ca0+n30_sly4_skoprime} 
Transition densities to the $0^+$ excited states of $^{50}$Ca in a small bump at low-energy (SLy4). 
}
\end{figure}

\begin{figure}
\includegraphics{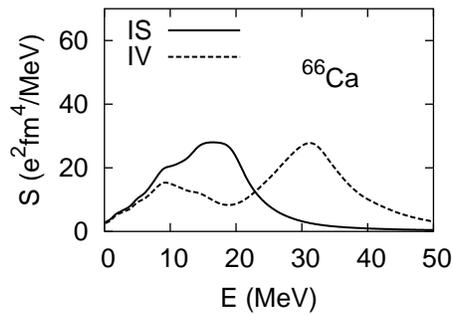}
\caption{\label{fig:str_ca0+n46_sly4} 
0$^+$ strength functions of $^{66}$Ca (SLy4).
}
\end{figure}

Next we compare predictions for $^{50}$Ca.
Figure \ref{fig:tr_den_ca0+n30_sly4_skoprime} shows the SLy4 transition
densities to the states in the small bump 
in the low-energy continuum.  The bump emerges there with both interactions. 
$^{66}$Ca is the neutron-drip-line nucleus obtained with SLy4; 
Figure \ref{fig:str_ca0+n46_sly4} displays the strength functions in that
nucleus.  
Apparently they are similar not to those of the
SkM$^{\ast}$ drip-line nucleus $^{76}$Ca but rather to 
those of the SKM$^{\ast}$ $^{66}$Ca (see
Fig.~\ref{fig:str_ca_0+}).  Clearly the quantum numbers of the quasiparticle
orbits involved in the transition affect the transition density much more than the
chemical potential.

\begin{figure}
\includegraphics{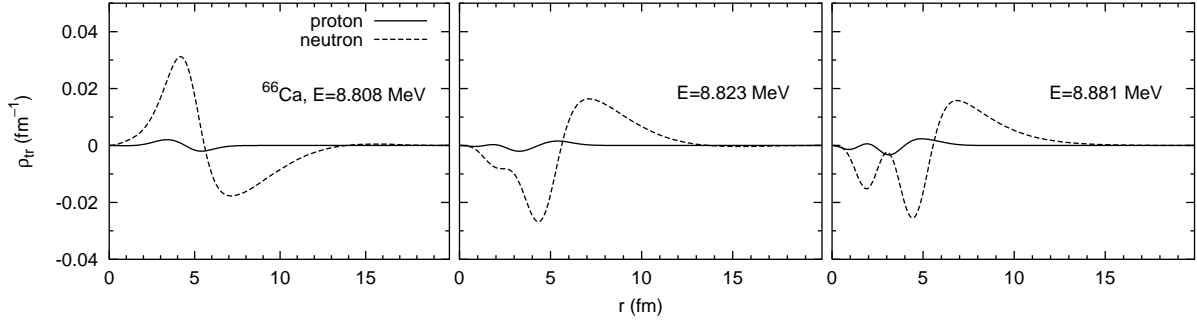}
\caption{\label{fig:tr_den_ca0+n46_sly4} 
Transition densities to the excited states in the 
low-energy shoulder (IS) or peak (IV) around $E=9$ MeV of the strength function  
of Fig.~\ref{fig:str_ca0+n46_sly4} (SLy4).}
\end{figure}
\begin{figure}
\includegraphics{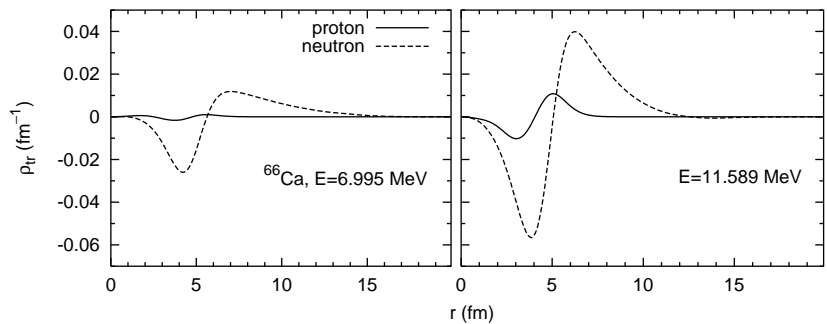}
\caption{\label{fig:tr_den_ca0+n46} 
Transition densities to the excited states in the 
low-energy shoulders around $E=9$ MeV and 11 MeV in the $0^+$ strength function  
of $^{66}$Ca (SkM$^\ast$).}
\end{figure}

 Figures 
\ref{fig:tr_den_ca0+n46_sly4} (SLy4) and 
\ref{fig:tr_den_ca0+n46} (SkM$^\ast$)  
show the transition densities 
to the excited states in the shoulder at about 9 MeV in $^{66}$Ca.
The positions of the outer neutron peak are nearly the same in both, 
but the neutron transition densities for $r<4$ fm have some differences.
The proton transition density is small except in the right panel of 
Fig.~\ref{fig:tr_den_ca0+n46}.
The tails of the transition densities are quite different 
from those in Fig.~\ref{fig:tr_den_ca0+n56}, for the nucleus with 10 more neutrons.
The two-quasiparticle components with the largest amplitude involve either
neutron $g_{9/2}$'s or $p_{3/2}$'s, depending on the state and interaction, and 
the largest amplitude itself ranges from 0.5 to 0.99.

\clearpage

\subsection{\label{subsec:transition density 1-} The 1$^{\bm -}$ channels}
 
 As we have just seen, the large neutron tails in the transition density are
responsible for enhanced low-lying strength near the drip line.  The robust enhancement of
low-lying IS $1^-$ strength is due in large part to the factor $r^3$, which
emphasizes large radii, in the
transition operator.

\begin{figure}
\includegraphics{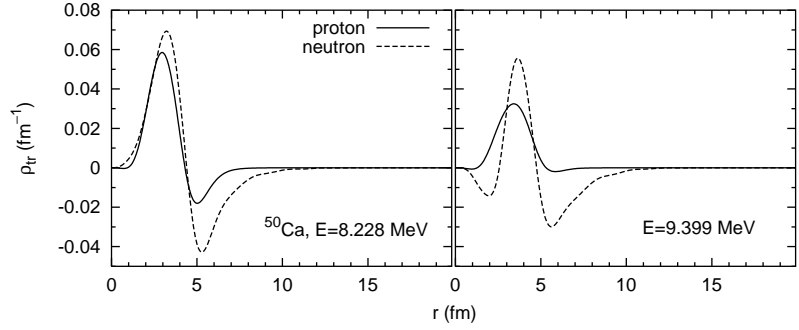}
\caption{\label{fig:tr_den_ca1-n30} 
Transition densities to the excited states in
the low-energy peak of the IS $1^-$ strength function of $^{50}$Ca.
 }
\end{figure}

One stark difference between these transition densities and those of
the $0^+$ channel
is the role played by protons, which is non-negligible here. Figure
\ref{fig:tr_den_ca1-n30} shows the transition densities of two states in the
low-lying IS peak in $^{50}$Ca.   Protons and neutrons contribute coherently.
The component of the state at $E=8.228$ MeV with the largest amplitude is 
$\nu 2 p_{3/2}({\cal E}=0.880$ MeV, $\bar{v}_\mu^2=0.488) \otimes
\nu 1 d_{3/2}({\cal E}=8.219$ MeV, $\bar{v}_\mu^2=0.981)$.  The squared amplitude 
$\bar{X}_{\mu\nu}^2-\bar{Y}_{\mu\nu}^2$ is only 0.36. 
Two other configurations,
$\nu 3s_{1/2}\otimes \nu 2 p_{3/2}$ and
$\pi 2p_{3/2}\otimes \pi 1 d_{3/2}$ have
$\bar{X}_{\mu\nu}^2-\bar{Y}_{\mu\nu}^2= 0.20$ and 0.12,  not much
smaller.  And since
the two-quasiparticle energy of the largest component is 9.099 MeV, 
the excited state gains more correlation energy than those 
in the $0^+$ channel.

\begin{figure}
\includegraphics{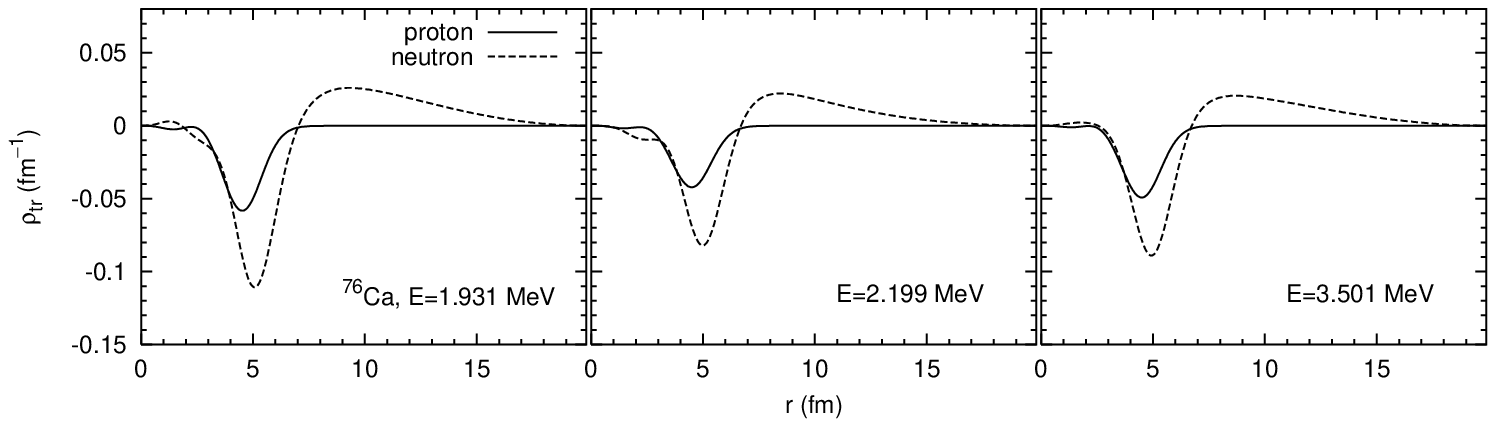}
\caption{\label{fig:tr_den_ca1-n56} 
Transition densities to the excited states in
the low-energy peak of the IS $1^-$ strength function of $^{76}$Ca.
 }
\end{figure}

The proton contributions remain even at the neutron drip line.
Figure \ref{fig:tr_den_ca1-n56} shows the transition densities
to 3 excited states in 
the large low-energy IS peak in
$^{76}$Ca (Fig.~\ref{fig:str_ca_1-}).
These states have IS strengths
9227 $e^2$fm$^6~(E=1.931$ MeV), 
3219 $e^2$fm$^6~ (E=2.199$ MeV), and
4856 $e^2$fm$^6~ (E=3.501$ MeV), 
representing 42 \%, 15 \%, and 22 \% of 
the total strength.
While protons play no role for $r>6$ fm, they are active around $r=5$ fm.
The main difference between these states  and those of $^{50}$Ca is that here
the neutrons both extend much further out and contribute more inside the
nucleus.  It is the tail that is mainly responsible for the increased strength.  

Interestingly, the states in $^{76}$Ca are mainly of two-quasiparticle
character even though there is a coherent proton contribution.  The component
in the state at $E=1.931$ MeV with the largest amplitude
($\bar{X}_{\mu\nu}^2-\bar{Y}_{\mu\nu}^2=0.914$) is
$\nu 3s_{1/2}({\cal E}=0.355$ MeV, $\bar{v}_\mu^2= 0.645)\otimes 
 \nu 3p_{3/2}({\cal E}=1.582$ MeV, $\bar{v}_\mu^2=0.592\times 10^{-3})$. 
In the state at $E=2.199$ MeV the component with the largest amplitude
($\bar{X}_{\mu\nu}^2-\bar{Y}_{\mu\nu}^2=0.975$) is 
$\nu 3p_{3/2}\otimes\nu 2d_{5/2}({\cal E}=0.618$ MeV, $\bar{v}_\mu^2= 0.776)$, 
and at $E=3.501$ MeV it is 
$\nu 2d_{5/2}\otimes
 \nu 2f_{7/2}({\cal E}=2.860$ MeV, $\bar{v}_\mu^2=0.1\times 10^{-3})$, with
$\bar{X}_{\mu\nu}^2-\bar{Y}_{\mu\nu}^2=0.936$.
The proton two-quasiparticle configuration with the largest amplitude 
has $\bar{X}_{\mu\nu}^2-\bar{Y}_{\mu\nu}^2=(0.2-0.3)\times 10^{-2}$.

\begin{figure}
\includegraphics{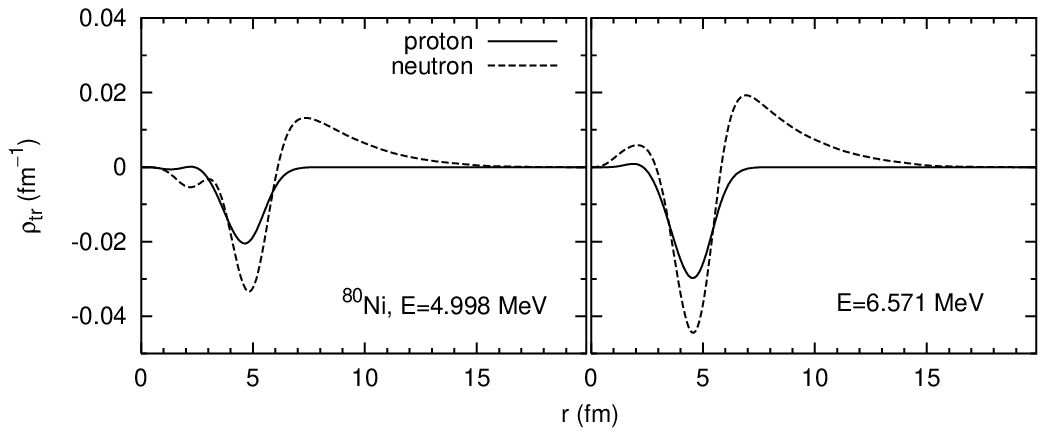}
\caption{\label{fig:tr_den_ni1-n52} 
Transition densities to the excited states in
the low-energy peak of the IS $1^-$ strength function of $^{80}$Ni.
 }
\end{figure}
\begin{figure}
\includegraphics{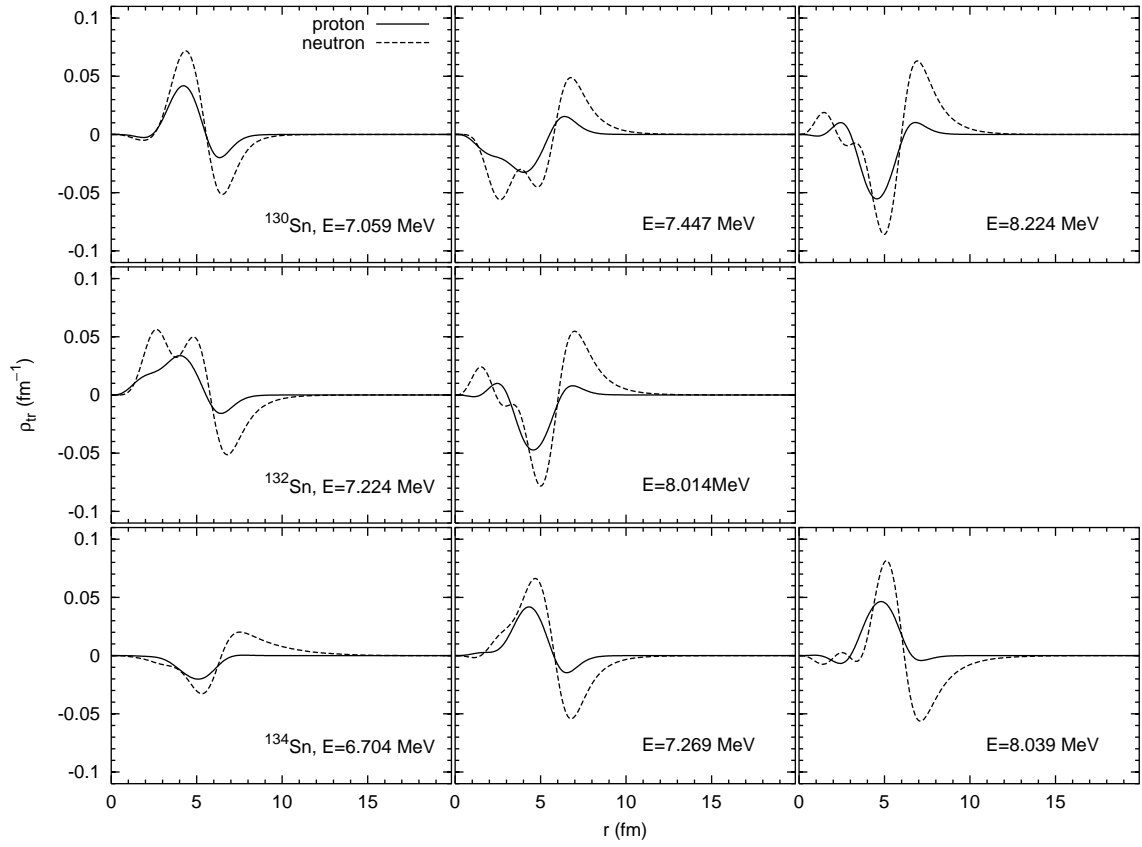}
\caption{\label{fig:tr_den_sn1-n80_n84} 
Transition densities to the excited states in
the low-energy peak of the IS $1^-$ strength function of
$^{130}$Sn (upper panels), $^{132}$Sn (middle), and $^{134}$Sn (lower).
 }
\end{figure}

In Ni, a low-energy peak rises suddenly at $^{80}$Ni
(Fig.~\ref{fig:str_ni_1-}).  The transition densities of the states in the
peak, shown in Fig.~\ref{fig:tr_den_ni1-n52}, are similar to those of the
neutron-rich Ca isotopes.  The density in $^{98}$Ni, at the drip line, is
similar though larger.
In Sn there is no sudden increase in the low-energy peak around $N=82$ (see
Fig.~\ref{fig:str_sn_1-}).  Fig.~\ref{fig:tr_den_sn1-n80_n84} shows the
transition densities to low-energy states in three isotopes around $N=82$.
Here the proton contributions stretch to larger radii than in Ca or Ni, even
though they are dwarfed by the neutron tails there.

The other Skyrme interactions give very similar results, both in the IS and IV
channels and for all three isotope chains.

\clearpage

\clearpage

\subsection{\label{subsec:transition_density_2+}The 2$^{\bm +}$ channels}
\begin{figure}
\includegraphics{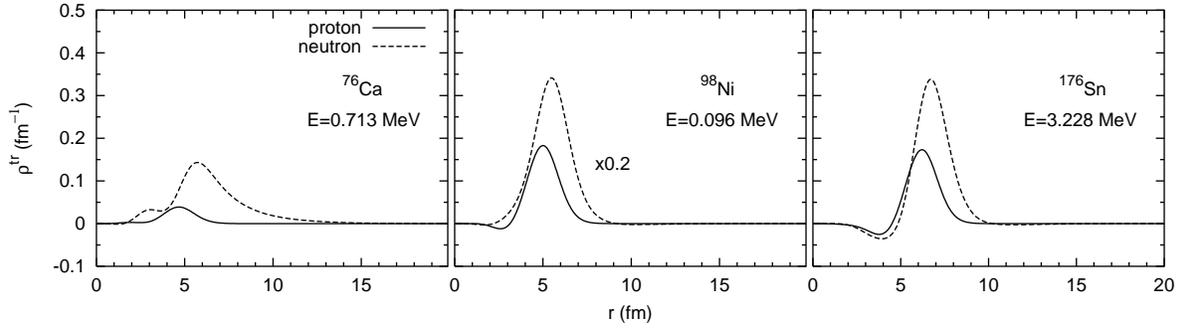}
\caption{\label{fig:tr_den_sn2+n126_ni2+n70_ca2+n56} 
Transition densities to the excited states in
the low-energy peak of the IS $2^+$ strength functions in 
the neutron-drip-line nuclei (wth SkM$^\ast$)
$^{76}$Ca, $^{98}$Ni, and $^{176}$Sn.
Both proton and neutron curves are reduced by a factor of 5 in $^{98}$Ni to
fit on the figure..
 }
\end{figure}

Figures \ref{fig:str_ca_2+}--\ref{fig:str_sn_2+}
showed that at the neutron drip line 
the IS and IV strength functions in Ni and Sn have distinct low-energy peaks
(though the
IS strength is always larger, even at the low energy IV peak) while in Ca the
peaks coincided.  This is true, except in Ni with SLy4\footnote{$^{88}$Ni has
a small IV peak at the IS peak.},
no matter which Skyrme interaction we use.
With SkM$^{\ast}$, the IS strength of the large low-energy peak 
at $E=0.096$ MeV in $^{98}$Ni is 92 \% of
the total strength (8.9 \% of the EWSR), and 
that of the peak at $E=0.713$ MeV in $^{176}$Sn is 23 \% of the total strength
(2.2 \% of the EWSR).   
In Fig.\ \ref{fig:tr_den_sn2+n126_ni2+n70_ca2+n56} we compare the transition
densities of the states in these low-energy peaks with those in $^{76}$Ca. 
As  Fig.~\ref{fig:str_ca_2+} suggests, though the protons play a
role, the neutrons are proportionately more important in Ca and do not
seem to correlate with
the protons.  In $^{98}$Ni and $^{176}$Sn, by contrast, the neutrons and protons are
completely in phase (as they must be if the IV strength is
negligible). 

The other Skyrme interactions bear out these conclusions, though in the 
SLy4 calculation of the drip-line Ca isotope the
ratio of proton transition density to  
neutron transition density is larger than that with SkM$^\ast$
(shown in the left 
panel of Fig.~\ref{fig:tr_den_sn2+n126_ni2+n70_ca2+n56}).  In all nuclei
near the neutron drip line, the states in low-energy peaks are mostly
above neutron-emission threshold, yet the neutron tails 
cut off at much smaller radii than to those in the $0^+$ and $1^-$ channels.
In addition, the transition densities in those other channels generally have
a node at about $r=5$ fm, while the $2^+$ transition densities show no such feature. 

\begin{figure}
\includegraphics{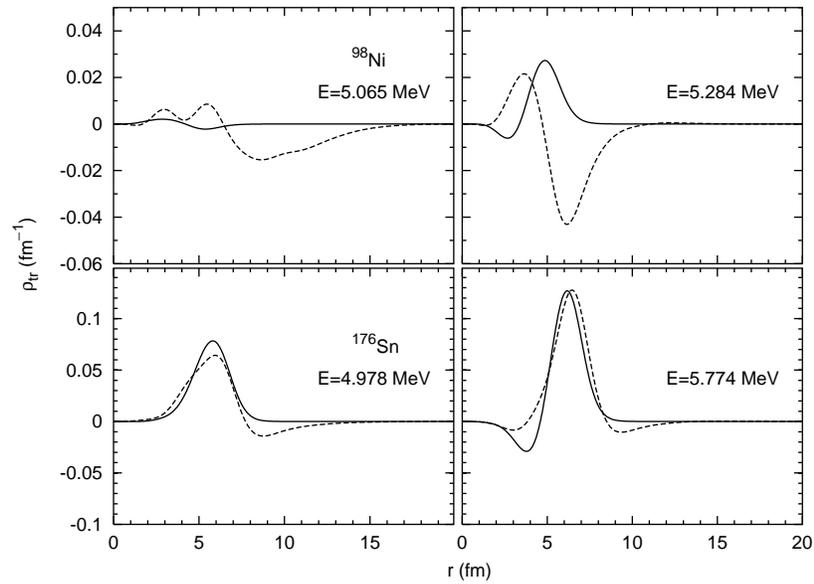}
\caption{\label{fig:tr_den_2_sn2+n126_ni2+n70} 
Transition densities 
to the excited states in
the low-energy peak of the IV $2^+$ strength functions
(SkM$^\ast$) in 
$^{176}$Sn and $^{98}$Ni.
}
\end{figure}

We put off for just a moment the obvious question of
whether these low-lying IS peaks are familiar quadrupole vibrations and turn
briefly to the IV channel.
Figure \ref{fig:tr_den_2_sn2+n126_ni2+n70} shows
the transition densities to states in 
the low-energy IV strength-function peak in 
$^{176}$Sn and $^{98}$Ni.
The states in $^{176}$Sn display remarkably IS character, despite
being at the peak
of the IV distribution.  The fact that the IV distribution peaks there
is a little misleading because the IS strength is still considerably larger,
even though it is larger still at lower energies.  We get some IV strength
because the ratio of the transition density of protons to that of neutrons is not
exactly equal        to $Z/N$ (see Eq.~(\ref{eq:transition_operator_2+})).

Returning to the IS strength, we note, in accord with Fig.\
\ref{fig:tr_den_sn2+n126_ni2+n70_ca2+n56}, that the transitions appear
more collective in the Ni and
Sn drip-line isotopes than in that of Ca.  The component with the largest
amplitude ($\bar{X}_{\mu\nu}^2-\bar{Y}_{\mu\nu}^2=0.81$)
in the state at $E=0.713$ MeV of $^{76}$Ca 
is $\nu 3 s_{1/2}({\cal E}=0.355$ MeV, $\bar{v}_\mu^2=0.65)
\otimes
\nu 2 d_{5/2}({\cal E}=0.618$ MeV, $\bar{v}_\mu^2=0.78)$, and
that with the second largest
($\bar{X}_{\mu\nu}^2-\bar{Y}_{\mu\nu}^2=0.16$)
is 
$\{\nu 2 d_{5/2}\}^2$.
This is still a predominantly single two-quasiparticle excitation.
By contrast, the components of the state with at $E=0.096$ MeV in $^{98}$Ni
with the largest amplitudes are 
\begin{eqnarray*}
&& \{\nu 1 h_{11/2}({\cal E}=1.415 \text{ MeV}, \bar{v}_\mu^2=0.27)\}^2, 
\bar{X}_{\mu\nu}^2-\bar{Y}_{\mu\nu}^2=0.35, \\
&& \{\nu 1 g_{7/2}({\cal E}=1.350 \text{ MeV}, \bar{v}_\mu^2=0.59)\}^2, 
\bar{X}_{\mu\nu}^2-\bar{Y}_{\mu\nu}^2=0.30, \\   
&& \pi 2 p_{3/2}({\cal E}=2.352 \text{ MeV}, \bar{v}_\mu^2=0)
\otimes
\pi 1 f_{7/2}({\cal E}=2.222 \text{ MeV}, \bar{v}_\mu^2=1), 
\bar{X}_{\mu\nu}^2-\bar{Y}_{\mu\nu}^2=0.12.
\end{eqnarray*}
While still not very collective, this state is a little more so than its counterpart
in Ca.
($^{98}$Ni may be in the spherical-deformed transition 
region.)

In $^{176}$Sn, the component of the state at $E=3.228$ MeV with the
largest amplitude 
is (Fig.~\ref{fig:tr_den_sn2+n126_ni2+n70_ca2+n56}) 
$\nu 2 g_{9/2}({\cal E}=1.477$ MeV, $\bar{v}_\mu^2=0)
\otimes
\nu 1 i_{13/2}({\cal E}=2.475$ MeV, $\bar{v}_\mu^2=1.00)$, 
($\bar{X}_{\mu\nu}^2-\bar{Y}_{\mu\nu}^2=0.79$)
and that with the next largest is 
$\pi 2 d_{5/2}({\cal E}=3.654$ MeV, $\bar{v}_\mu^2=0)
\otimes
\pi 1 g_{9/2}({\cal E}=2.276$ MeV, $\bar{v}_\mu^2=1)$, 
($\bar{X}_{\mu\nu}^2-\bar{Y}_{\mu\nu}^2=0.09$).
This excitation is mainly a single neutron particle and neutron hole.
Still, the proton contribution is much larger 
than that in other channels at the neutron drip line.
This excited state has 23.7 \% of the total IS strength and  
8.6 \% of the EWSR.

\begin{figure}
\includegraphics[width=1.0\textwidth]{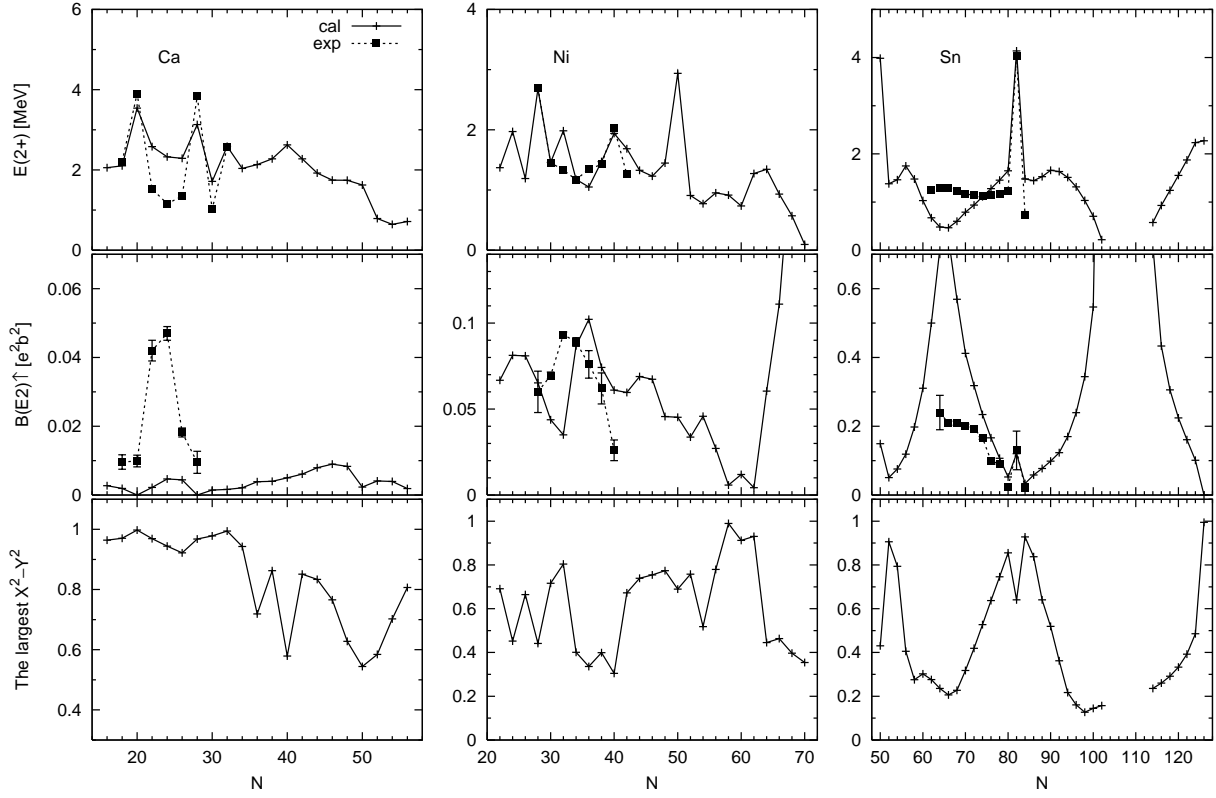}
\caption{\label{fig:n-e-be2-largest_sn_ni_ca2+} 
$N$-dependence of the energy (upper),  $B(E2;0^+\rightarrow 2^+)$ 
(middle), and largest $\bar{X}_{\mu\nu}^2-\bar{Y}_{\mu\nu}^2$
(lower) of the lowest $2^+$ states in Ca, Ni, and Sn, 
with SkM$^\ast$.  Experimental data are also shown where available.
}
\end{figure}

We see both in this channel and others that the large transition strength
near the neutron drip line 
does not always indicate the coherent contributions of many two-quasiparticle 
excitations (i.e., collectivity). Yet the transition densities make these
strong low-lying states, particularly in Ni and Sn, look very much like
the surface vibrations that characterize nuclei closer to stability.
To get a better handle on the degree to which the states we examine here can be called
vibrations, we show in Fig.~\ref{fig:n-e-be2-largest_sn_ni_ca2+}
the energy, transition 
probability $B(E2;0^+\rightarrow 2^+)$, and largest
$\bar{X}_{\mu\nu}^2-\bar{Y}_{\mu\nu}^2$ for neutrons in the lowest-energy $2^+$ states of 
Ca, Ni, and Sn.
The Ca panel shows, first of all, that we do not reproduce experimental data
very well except in a few nuclei, and second, that the states are
dominated by a single two-quasiparticle excitation for $N\leq 34$.
For $34 \leq N \leq 40$ something peculiar happens; when the energy increases,
so does the $B(E2;0^+\rightarrow 2^+)$, and vice versa, with the
largest $\bar{X}_{\mu\nu}^2-\bar{Y}_{\mu\nu}^2$, a measure of collectivity,
varying irregularly.
In nuclei with collective vibrations, transition strength is usually correlated
inversely with energy \cite{Gro62,Ram01}; we see the opposite here.  
Near the neutron drip line of Ni, the classic signs of
increasing quadrupole collectivity  --- decreasing energy, rising
$B(E2;0^+\rightarrow 2^+)$, and decreasing maximum
$\bar{X}_{\mu\nu}^2-\bar{Y}_{\mu\nu}^2$ --- are all present.
The Sn isotopes look much more familiar, exhibiting the usual trends, except 
at the magic numbers.  And the states close to the drip line with $96<N<124$
are as classically collective as (or more so than) those in more stable isotopes
with $56<N<68$.

\begin{figure}
\includegraphics{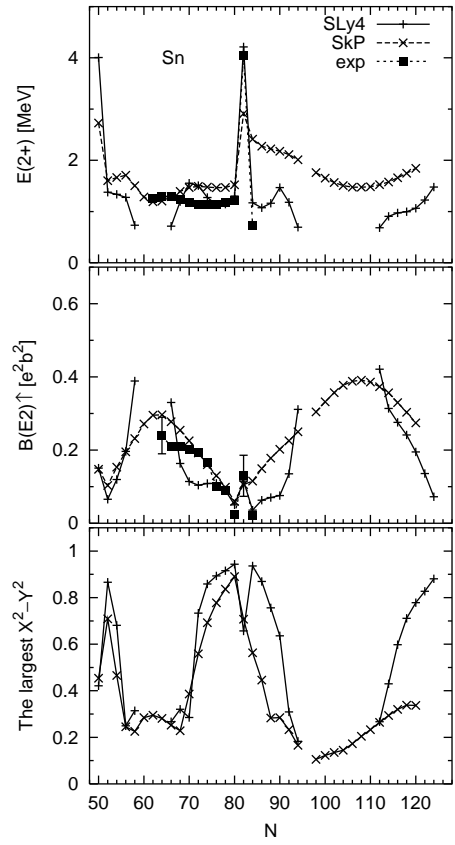}
\caption{\label{fig:n-e-be2-largest_sly4_skp_skoprime} 
The same as Fig.~\ref{fig:n-e-be2-largest_sn_ni_ca2+}
but for Sn only, SLy4 and SkP.   
}
\end{figure}

Finally we show the interaction dependence of the collectivity
measures for the lowest $2^+$ states in Sn (the easiest chain to understand)
in Fig.~\ref{fig:n-e-be2-largest_sly4_skp_skoprime}.
In the mid-shell region, 
the energies are low, the $B(E2;0^+\rightarrow 2^+)$'s are high, and 
the largest amplitudes low, as expected from systematics in stable nuclei in
regions of large dynamical deformation.  The values of these quantities depend
significantly on the interaction, however.  Ca and Ni show similar behavior but
with more irregularities.
\clearpage

\clearpage

\section{\label{sec:conclusion}Conclusion}

The feature that leaps out of our calculations is the enhancement of
low-energy strength  as the neutron drip line is
approached.  It is present in all multipoles (and is most pronounced in the
IS $1^-$ channel, where the transition operator contains a factor $r^3$), with all Skyrme interactions,
and in every isotopic chain, beginning just after a closed neutron shell.
Interestingly, the recent measurement of an IV $1^-$ 
strength function at GSI shows a low-energy peak before a closed shell, 
in $^{130}$Sn.  This may be due to correlations not included in the QRPA.

The transition densities to the low-lying states reveal some interesting
features. In the $0^+$ and $1^-$ channels, the low-energy peaks are usually
created essentially entirely by a single two-neutron-quasiparticle configuration.  In the $0^+$ channel, where the total number of two-quasiparticle
configurations is restricted, even small admixtures of two-proton-quasiparticle
states are absent; the proton transition density is nearly zero.   Protons play
a noticeably larger
role in the transition density of the $1^-$ channel despite the lack of
any two-proton-quasiparticle
configurations with sizable amplitude. 
In both these channels, the transition densities have a peak around the rms
radius and very long tail outside the radius, which allows single quasiparticle
configuration to have such large strength.

The transition densities to $2^+$ states are different; they have no real nodes
and a proton component that is of the same order as the neutron component.
The transition densities near the neutron-drip line resemble those of
surface vibrations, and in many cases the
corresponding states clearly are vibrational.  But the usual systematics of
collective vibrations is not always present; energies and $E$2 strengths are
sometimes correlated in the wrong way.  

Some of the most significant differences in the predictions of the three Skyrme
interactions we used are in static properties:  the locations of the neutron
drip line and quadrupole-deformed regions.  But some dynamical properties also
depend significantly on the interaction, most notably (and easiest to measure)
the energies of and transition-strengths to the lowest $2^+$ states near
regions of deformation and the total energy-weighted IV strength.  

This paper, in a way, is a catalog of predicted
dynamic properties of nuclei near the
drip line.  For the most part we have not offered simple explanations for the
anomalies that appear.  That, we hope, will be the subject of a future paper.

\acknowledgments
This work was supported in part by the U.S. Department of Energy under grant
DE-FG02-97ER41019.  We used computers belonging to the National Center for Computational 
Sciences at Oak Ridge National Laboratory and 
Information Technology Services at University of North 
Carolina at Chapel Hill.
Parts of this research were done when one of us
(J.T.) was at the University of North Carolina at Chapel Hill 
and at RIKEN.
\appendix*
\section{\label{app:tr_den} transition density}
The transition density is defined by
\begin{equation}
\rho_{\rm tr}^q(\bm{r};k) = \langle k| \sum_{i=1}^{N {\rm or} Z}
\delta(\bm{r}-\bm{r}_i) 
 |0\rangle, 
\end{equation}
where $|0\rangle$ and $|k\rangle$ denote the ground and excited states 
of the system, and $\bm{r}_i$ in the right-hand side is an operator
acting on the nucleon $i$. The transition density is evaluated in the QRPA 
(we omit $q$) as
\begin{eqnarray}
\rho_{\rm tr}(\bm{r};k) &=& 
\sum_{K<K^\prime}\left\{
X^{k\ast}_{KK^\prime}\left( \psi_{K^\prime}^\dagger(\bm{r})
\psi_{\bar{K}}(\bm{r}) u_{K^\prime} v_K 
-\psi_K^\dagger(\bm{r})\psi_{\bar{K}^\prime}(\bm{r})u_K v_{K^\prime}
\right) \right.\nonumber \\
&& \left.
-Y_{KK^\prime}^{k\ast}\left( \psi_{\bar{K}^\prime}^\dagger(\bm{r})
\psi_K(\bm{r}) v_{K^\prime} u_K
-\psi_{\bar{K}}^\dagger \psi_{K^\prime}(\bm{r}) v_K u_{K^\prime}
\right)
\right\}, \label{eq:tr_den}
\end{eqnarray}
where $X_{KK^\prime}^{k}$ and $Y_{KK^\prime}^{k}$ are the forward and 
backward amplitudes of the QRPA solution $k$ (see Appendix A of
Ref.~\cite{Ter05}), and $\psi_K(\bm{r})$ 
stands for a single-particle canonical-basis wave function.
The $uv$ factors associated with the basis are represented by
$u_K$ and $v_K$.
Assuming spherical symmetry for the system and using 
\begin{equation}
\psi_K(\bm{r}) = \sum_{m\sigma}R_\mu(r)Y_{ll_z}(\Omega)
\langle ll_z\frac{1}{2}\sigma|jm\rangle |\sigma\rangle,
\end{equation}
with $K=(\mu m)=(nljm)$ a set of quantum numbers,  
$R_\mu(r)$ a radial wave function, and $|\sigma\rangle$ a spin wave function, 
one can derive  
\begin{eqnarray}
\rho_{\rm tr}(\bm{r};k) &=& 
\sum_{\mu\leq\mu^\prime}\frac{1}{\sqrt{1+\delta_{\mu\mu^\prime}}}
\sqrt{(2j+1)(2j^\prime+1)}
R_{\mu^\prime}(r) R_\mu(r) \nonumber\\
&&\times(-)^{j+\frac{1}{2}}
\frac{1}{\sqrt{4\pi}}
\left(\begin{array}{ccc}j^\prime & j & J_k \\
-1/2 & 1/2 &0 \end{array}\right)
\frac{1}{2}(1+(-)^{l^\prime+l+J_k})
Y_{J_kM_k}^\ast(\Omega) \nonumber \\
&& \times((-)^{J_k}v_{\mu^\prime}u_\mu 
+v_\mu u_{\mu^\prime})
(\bar{Y}^k_{\mu\mu^\prime}+(-)^{J_k}\bar{X}^k_{\mu\mu^\prime}),
\end{eqnarray}
where $\mu^\prime=(n^\prime j^\prime l^\prime)$, and 
$J_k$ denotes the angular momentum of the state $k$. 
$\bar{X}_{\mu\mu^\prime}^k$ and 
$\bar{Y}_{\mu\mu^\prime}^k$ are the forward and backward 
amplitudes in the spherical representation.


\end{document}